\documentclass[aps,pra,twocolumn,groupedaddress,floatfix]{revtex4}

\usepackage{amsmath}
\usepackage{graphicx}
\usepackage{multirow}
\usepackage{color}

\begin{document}

\title{Generalized gradient expansion for inhomogeneous dynamical mean-field theory: Application to ultracold atoms in a harmonic trap}
\author{J. K. Freericks, Shuyang Han, and Karlis Mikelsons}
\affiliation{Department of Physics, Georgetown University, 37$^{\rm th}$ and O Sts. NW, Washington, DC 20057, USA}
\author{H. R. Krishnamurthy}
\affiliation{Department of Physics, Indian Institute of Science, Bangalore-560012, India}

\date{\today}

\begin{abstract}
We develop a generalized gradient expansion of the inhomogeneous dynamical mean-field theory method
for determining properties of ultracold atoms in a trap. This approach goes beyond the well-known local density approximation and at higher temperatures, in the normal phase, it shows why the local density approximation works so well, since the local density and generalized gradient approximations are essentially indistinguishable from each other (and from the exact solution within full inhomogeneous dynamical mean-field theory). But because the generalized gradient expansion only involves nearest-neighbor corrections, it does not work as well at low temperatures, when the systems enter into ordered phases.  This is primarily due to the problem that ordered phases often satisfy some global constraints which determine the spatial ordering pattern, and the local density and generalized gradient approximations are not able to impose
those kinds of constraints; they also overestimate the tendency to order.   The theory is applied to phase separation of different mass fermionic mixtures represented by the Falicov-Kimball model and to determining the entropy per particle of a fermionic system represented by the Hubbard model. The generalized gradient approximation is a useful diagnostic for the accuracy of the local density approximation---when both methods agree, they are likely accurate, when they disagree, neither is likely to be correct.
\end{abstract}

\pacs{to be determined}
\date{\today}

\maketitle

\section{Introduction} 

Ultracold atoms in optical lattices have become one of the most interesting platforms to examine many-body
physics properties of quantum particles. Experiments are under great control, and one can modify both lattice 
parameters and interactions by adjusting laser intensity or external magnetic field strength. One of the challenges, however, is that these systems
also require a (harmonic) trap to keep them stable, which breaks the periodicity of the system. The system size is also finite---usually about $100-300$ lattice sites per spatial dimension---implying a three-dimensional system has one to thirty million lattice sites. Many different numerical techniques have been applied to these systems, but the larger systems in two-dimensions and nearly all systems in three-dimensions are too large for most exact calculations (the one exception is Bosonic systems, where the quantum-Monte-Carlo approaches based on the worm algorithm work extremely well in
equilibrium for $\approx$~300,000 particles~\cite{prokofiev_qmc1,prokofiev_qmc2}). Inhomogeneous dynamical mean-field theory (IDMFT) has been applied to many of these systems, and it works well at high temperatures, but it often cannot handle the full system size, especially in three-dimensions~\cite{rosch1,blumer}.  The dynamical mean-field theory (DMFT) has two computationally limiting steps---one is calculating the diagonal of the inverse of a general complex matrix whose dimension is the number of lattice sites $|\Lambda|$ and the other is solving the impurity problem on each lattice site. The former problem grows like $|\Lambda|^3$ for conventional linear-algebra algorithms, while the latter grows linearly with $|\Lambda|$.  Hence, significant progress can be made with the IDMFT approach if other methods could be employed to approximately solve the calculation of the local Green's function from the local self-energy. Other, more direct, quantum Monte Carlo approaches, also suffer from similar issues regarding the computational effort required (which usually grows as a power law of $|\Lambda|$) and so are also limited by the system size they can work with~\cite{scalettar}.

This problem has been tackled by employing the so-called local density approximation (LDA), which assumes the local properties of a site within a trap
at a given local chemical potential are well approximated by the local properties of a site inside a periodic system at the same chemical potential~\cite{kohn_sham,lda_atoms}. Since the trap varies both smoothly and slowly with position, this approximation ends up being quite good, with the chemical potential adjusted 
for each temperature to conserve the total number of particles. One expects it to start to fail in two regimes: (i) the
first is at low temperature for ordered phases, especially ordered phases which partially break translational symmetry, because the LDA does not invoke any global constraints involved with ensuring that the ordered phase can ``fit'' into the given lattice structure with the trap present and (ii) the second is in nonequilibrium systems with mass transport, because the LDA
assumes the density at a given site is unchanged regardless of the external field applied. These two issues  motivate us to try and develop a technique that will share the success of the LDA for high temperatures but will be able to correct it as the temperature is lowered to take into account some of these nonlocal effects
or can handle mass transport in the presence of external perturbations (see Ref.~\onlinecite{dirks} for an application of LDA in nonequilibrium when there is no mass transport).

A similar problem has existed within band structure calculations for quite some time. The generalized gradient approximation (GGA) was introduced there to try to take into account some of this nonlocal behavior~\cite{gga}. One can think of it, in some sense, as a Taylor series expansion in ``inhomogeneity'' with the LDA as the lowest-order term, and gradient corrections as the next term in the hierarchy. In this work, we extend this GGA concept to solving for the Green's functions of 
inhomogeneous systems that {\it takes into account the local inhomogeneity about a given lattice site}. While the initial formulas we derive will hold for all computations, we quickly adopt them to the DMFT approach, where we assume the self-energy is local. Another way of describing this calculation
is that both the LDA and the GGA approximate the step where we calculate the diagonal of the inverse of a $|\Lambda|\times|\Lambda|$ matrix; the first by replacing the diagonal in the inverse matrix by one matrix element for the whole diagonal and using Fourier transformation to compute the inverse (LDA) and the second by setting up a self-consistent equation for the diagonal of the inverse that involves employing the LDA local and nearest-neighbor Green's functions and the local self-energy and GGA Green's function (GGA). Other forms of extending the LDA to a GGA have also been proposed. For example, see Ref.~\onlinecite{gga-other}.

One can view this work as providing a rigorous expansion about the LDA and demonstrating why it works so well in the normal state. It also provides, in principle, a path toward systematically improving the LDA, but, as we show below, it is unlikely one will achieve high accuracy at low temperatures with any variant of this approach. Hence, other approaches, which take into account the sparsity of the system, are likely to be required for even more accurate solutions of these problems. But, when the LDA and GGA agree, this is a strong indication that the LDA is accurate for this set of parameters, and this can be a quite useful application of the GGA approach.

The formalism is developed in Sec. II for the two models we consider in this work: the spinless fermionic Falicov-Kimball model and the fermionic Hubbard model. In Sec. III, we present the numerical results
for ordered phases in mixtures and for the entropy in the Hubbard model. Conclusions and outlooks are given in Sec. IV.

\section{Formalism}

In order to determine the properties of a many-body quantum system, we often employ a Green's function-based method. The imaginary-time Green's function is defined by
\begin{equation}
G_{i,j;\sigma}(\tau)=-{\rm Tr}\left \{ \frac{e^{-\beta\mathcal{H}}}{\mathcal{Z}}\mathcal{T}_\tau
c_{i\sigma}^{\phantom{\dagger}}(\tau)c^\dagger_{j\sigma}(0)\right \}.
\label{eq: g_tau}
\end{equation}
Here, $\mathcal{H}$ is the Hamiltonian, $\mathcal{Z}={\rm Tr}\exp[-\beta\mathcal{H}]$ is the partition function, $\beta=1/T$ is the inverse temperature, $-\beta\le \tau\le\beta$ is the imaginary time and $\mathcal{T}_\tau$ is the time-ordering operator, indicating that the operators are ordered from right to left in increasing time. The fermionic
creation (annihilation) operators are $c^\dagger_{i\sigma}$ ($c^{\phantom{\dagger}}_{i\sigma}$)
and $\mathcal{O}(\tau)$ denotes an operator in the Heisenberg representation $\mathcal{O}(\tau)=
\exp[\mathcal{H}\tau]\mathcal{O}\exp[-\mathcal{H}\tau]$. The index $i$ ($j$) is the lattice site and $\sigma$ is the $z$-component of the spin of the fermion (which is dropped for the spinless Falicov-Kimball model). We will show the explicit Hamiltonians below.

We also can define the Green's function for real times, and here, we will define the retarded Green's function, which satisfies
\begin{equation}
G_{i,j;\sigma}^R(t)=-i\theta(t){\rm Tr}\left \{ \frac{e^{-\beta\mathcal{H}}}{\mathcal{Z}}
\{c_{i\sigma}^{\phantom{\dagger}}(t)c^\dagger_{j\sigma}(0)\}_+\right \},
\end{equation}
where $\theta(t)$ is the Heaviside unit-step function, $\{\ldots,\ldots\}_+$ denotes the anticommutator,
and the Heisenberg representation for an operator in real time is $\mathcal{O}(t)=
\exp[i\mathcal{H}t]\mathcal{O}\exp[-i\mathcal{H}t]$. We typically work with the Fourier transform of the
retarded Green's function, which is
\begin{equation}
G_{i,j;\sigma}^R(\omega)=\int_{0}^{\infty} dt e^{i\omega t} G_{i,j;\sigma}^R(t).
\end{equation}

The derivation of the GGA formalism starts with the Dyson equation for the retarded Green's function with the inhomogeneous Hamiltonian on a lattice $\Lambda$ (with $|\Lambda|$ lattice sites), which includes a (global) chemical potential ($\mu$), a local potential for lattice site $i$ ($V_i$ with $\mu_i=\mu-V_i$ the local chemical potential), a hopping between lattice sites $i$ and $j$ for spin $\sigma$ ($-t_{i,j;\sigma}$), and a self-energy
corresponding to sites $i$ and $j$ [$\Sigma_{i,j;\sigma}^R(\omega)$]:
\begin{equation}
\sum_{k=1}^{|\Lambda|}\left [ (\omega+\mu_i)\delta_{i,k}+t_{i,k;\sigma}-\Sigma_{i,k;\sigma}^R(\omega)\right ] G_{k,j;\sigma}^R(\omega)=\delta_{i,j},
\label{eq: dyson1}
\end{equation}
which has an obvious modification on the imaginary axis when Matsubara frequencies are used instead of real frequencies.
We also want to investigate the corresponding equation in the bulk, which would be used for the LDA approach. To do this, we need to introduce some notation. For simplicity, we will assume the hopping matrix is homogeneous (this is not required, but makes the notation much less complex), so that $-t_{i,j;\sigma}$ is the same for the inhomogeneous system and for the homogeneous ones used to develop the LDA (for simplicity, we also assume the hopping is between nearest neighbors only, hence $-t_{i,j;\sigma}=-t_{i,i+\delta;\sigma}=-t_\sigma$ for $\delta$ a nearest-neighbor translation and $i+\delta$ a schematic notation indicating the neighbor of site $i$ in the $\delta$ direction). We then denote the (local) chemical potential for the homogeneous system by $\bar\mu_j^{(I)}=\mu-V_I$ for all $j\in\Lambda$. To be clear, here the index $j$ is the lattice site, but $\bar\mu_j^{(I)}$ doesn't actually depend on the lattice site $j$; we have set it globally equal to the value the inhomogeneous problem has at site $I$. If we write out the Dyson equation for the LDA Green's function corresponding to site $I$, it satisfies
\begin{eqnarray}
&~&\sum_{k=1}^{|\Lambda|}\left [ (\omega+\bar\mu_i^{(I)})\delta_{i,k}+t_{i,k;\sigma}-\Sigma_{i,k;\sigma}^{LDA(I)}(\omega)\right ] G_{k,j;\sigma}^{LDA(I)}(\omega)\nonumber\\
&~&=\delta_{i,j},
\label{eq: dyson_lda}
\end{eqnarray}
where we suppress the $R$ superscript for retarded. This system is actually translationally invariant and can be solved exactly in momentum space (and then Fourier transformed back to real space).

Since the right hand sides of Eqs.~(\ref{eq: dyson1}) and (\ref{eq: dyson_lda}) are equal, so are the corresponding left hand sides. First, we pick $i=j=I$ and solve for the local Green's function (by equating the corresponding left hand sides) to find
\begin{eqnarray}
&~&G_{I,I;\sigma}(\omega)=\frac{\omega+\bar\mu_I^{(I)}-\Sigma_{I,I;\sigma}^{LDA(I)}(\omega)}{\omega+\mu_I-\Sigma_{I,I;\sigma}(\omega)}G_{I,I;\sigma}^{LDA(I)}(\omega)\nonumber\\
&+&\frac{1}{\omega+\mu_I-\Sigma_{I,I;\sigma}(\omega)}\label{eq: gga_dyson}\\
&\times&\Biggr \{
\sum_{\delta}t_{I,I+\delta;\sigma}\left [
G_{I+\delta ,I;\sigma}^{LDA(I)}(\omega)-G_{I+\delta ,I;\sigma}(\omega)\right ]\nonumber\\
&+&\sum_{k\ne I}\left [ \Sigma_{I,k;\sigma}(\omega)G_{k,I;\sigma}(\omega)
-\Sigma_{I,k;\sigma}^{LDA(I)}(\omega)G_{k,I;\sigma}^{LDA(I)}(\omega)\right ] \Biggr \}.\nonumber
\end{eqnarray}
Here, we used the notation $\delta$ to denote a nearest-neighbor translation vector and $I+\delta$ to denote the lattice site corresponding to the translation by the nearest-neighbor translation vector $\delta$ from site $I$. For the next equation, we choose $i=I+\delta$, $j=I$ and the LDA is evaluated with respect to site $I+\delta$. We then solve for $G_{I+\delta ,I;\sigma}(\omega)$, which yields
\begin{widetext}
\begin{eqnarray}
&~&G_{I+\delta ,I;\sigma}(\omega)=\frac{\omega+\bar\mu_{I+\delta}^{(I+\delta)}-\Sigma_{I+\delta ,I+\delta;\sigma}^{LDA(I+\delta)}(\omega)}
{\omega+\mu_{I+\delta}-\Sigma_{I+\delta ,I+\delta;\sigma}(\omega)}G_{I+\delta ,I;\sigma}^{LDA(I+\delta)}(\omega)
+\frac{1}{\omega+\mu_{I+\delta}-\Sigma_{I+\delta ,I+\delta;\sigma}(\omega)}\label{eq:_dyson_offdiag}\\
&\times&\Bigr \{ \sum_{\delta^\prime}t_{I+\delta ,I+\delta+\delta^\prime;\sigma}\left [ G_{I+\delta +\delta^\prime ,I;\sigma}^{LDA(I+\delta)}(\omega)-
G_{I+\delta+\delta^\prime ,I;\sigma}(\omega)\right ]
+\sum_{k\ne I+\delta}\left [ \Sigma_{I+\delta ,k;\sigma}(\omega)G_{k,I;\sigma}(\omega)-
\Sigma_{I+\delta ,k;\sigma}^{LDA(I+\delta)}(\omega)G_{k,I;\sigma}^{LDA(I+\delta)}(\omega)\right ] \Bigr \}.\nonumber
\end{eqnarray}
For the GGA derivation, we want to restrict the terms in our self-consistent equations to involve only
on-site terms and nearest-neighbor terms, so we drop terms from Eq.~(\ref{eq:_dyson_offdiag}) that involve second neighbors or further. This implies $\delta^\prime=-\delta$ and $k=I$ only, which gives
\begin{eqnarray}
&~&G_{I+\delta ,I;\sigma}(\omega)=\frac{\omega+\bar\mu_{I+\delta}^{(I+\delta)}-\Sigma_{I+\delta ,I+\delta;\sigma}^{LDA(I+\delta)}(\omega)}
{\omega+\mu_{I+\delta}-\Sigma_{I+\delta ,I+\delta;\sigma}(\omega)}G_{I+\delta ,I;\sigma}^{LDA(I+\delta)}(\omega)
+\frac{1}{\omega+\mu_{I+\delta}-\Sigma_{I+\delta ,I+\delta;\sigma}(\omega)}\label{eq: dyson_offdiag2}\\
&\times&\Bigr \{ t_{I+\delta ,I;\sigma}\left [ G_{I,I;\sigma}^{LDA(I+\delta)}(\omega)-
G_{I,I;\sigma}(\omega)\right ]
+ \Sigma_{I+\delta ,I;\sigma}(\omega)G_{I,I;\sigma}(\omega)-
\Sigma_{I+\delta ,I;\sigma}^{LDA(I+\delta)}(\omega)G_{I,I;\sigma}^{LDA(I+\delta)}(\omega)\Bigr \}.\nonumber
\end{eqnarray}
Restricting $k=I+\delta$ in Eq.~(\ref{eq: gga_dyson}), and substituting in Eq.~(\ref{eq: dyson_offdiag2}),
produces (after some significant algebra and suppressing retarded superscripts and the $\sigma$ subscript)
\begin{eqnarray}
&~&G_{I,I}(\omega)=\frac{\omega+\mu_I-\Sigma_{I,I}^{LDA(I)}(\omega)}{\omega+\mu_I-\Sigma_{I,I}(\omega)}G_{I,I}^{LDA(I)}(\omega)
+\frac{1}{\omega+\mu_I-\Sigma_{I,I}(\omega)-\sum_\delta \frac{[t_{I,I+\delta}+\Sigma_{I,I+\delta}(\omega)][t_{I+\delta ,I}-\Sigma_{I+\delta ,I}(\omega)]}{\omega+\mu_{I+\delta}-\Sigma_{I+\delta ,I+\delta}(\omega)}}\nonumber\\
&\times&\Biggr \{ \sum_\delta t_{I,I+\delta}\left [ G_{I+\delta ,I}^{LDA(I)}(\omega)-\frac{\omega+\mu_{I+\delta}-\Sigma_{I+\delta ,I+\delta}^{LDA(I+\delta)}(\omega)}
{\omega+\mu_{I+\delta}-\Sigma_{I+\delta ,I+\delta}(\omega)}G_{I+\delta ,I}^{LDA(I)}(\omega)\right ]\nonumber\\
&~&-\sum_\delta\frac{\left [ t_{I,I+\delta}+\Sigma_{I,I+\delta}(\omega)\right ] \left [ t_{I+\delta ,I}-\Sigma_{I+\delta ,I}^{LDA(I+\delta)}(\omega)\right ]}{\omega+\mu_{I+\delta}-\Sigma_{I+\delta ,I+\delta}(\omega)} \left [ G_{I,I}^{LDA(I+\delta)}(\omega)-\frac{\omega+\mu_I-\Sigma_{I,I}^{LDA(I)}(\omega)}{\omega+\mu_I-\Sigma_{I,I}(\omega)}G_{I,I}^{LDA(I)}(\omega)\right ]\nonumber\\
&~&+\sum_\delta \left [ \Sigma_{I+\delta ,I}^{LDA(I+\delta)}(\omega)G_{I+\delta ,I}^{LDA(I+\delta)}(\omega)-\frac{\omega+\mu_I-\Sigma_{I,I}^{LDA(I)}(\omega)}{\omega+\mu_I-\Sigma_{I,I}(\omega)}\Sigma_{I+\delta ,I}(\omega)G_{I+\delta ,I}^{LDA(I)}(\omega)\right ]\nonumber\\
&~&+\sum_\delta \frac{\left [ t_{I,I+\delta}+\Sigma_{I,I+\delta}(\omega)\right ]\left [ \Sigma_{I+\delta ,I}^{LDA(I+\delta)}(\omega)-\Sigma_{I+\delta ,I}(\omega)\right ]}{\omega+\mu_{I+\delta}-\Sigma_{I+\delta ,I+\delta}(\omega)}\frac{\omega+\mu_I-\Sigma_{I,I}^{LDA(I)}(\omega)}{\omega+\mu_I-\Sigma_{I,I}(\omega)}G_{I,I}^{LDA(I)}(\omega)\Biggr \}.
\end{eqnarray}
\end{widetext}

In particular, if we examine this equation for a spatially uniform system, we have that the LDA self-energy equals the full self-energy, and the equation reduces to $G$=$G^{LDA}$, as it must.

Note that this result depends both on the local and the nearest-neighbor self-energy. If there isn't a method to be used to determine the nearest-neighbor self-energy of the inhomogeneous system, then one could use the ansatz
\begin{equation}
\Sigma_{I,J}(\omega)=\frac12 \left [ \Sigma_{I,J}^{LDA(I)}(\omega)+\Sigma_{I,J}^{LDA(J)}(\omega)\right ],
\end{equation}
to determine an approximation for that self-energy in terms of the LDA result. For this work, we focus on the IDMFT approach, so we have only local self-energies, and hence we have
$\Sigma_{I,J}(\omega)=0$. This result produces a great simplification, and we find
\begin{widetext}
\begin{eqnarray}
&~&G_{I,I}(\omega)=\frac{\omega+\mu_I-\Sigma_{I,I}^{LDA(I)}(\omega)}{\omega+\mu_I-\Sigma_{I,I}(\omega)}G_{I,I}^{LDA(I)}(\omega)
+\frac{1}{\omega+\mu_I-\Sigma_{I,I}(\omega)-\sum_\delta \frac{t_{I,I+\delta}t_{I+\delta ,I}}{\omega+\mu_{I+\delta}-\Sigma_{I+\delta ,I+\delta}(\omega)}}\nonumber\\
&\times&\Biggr \{ \sum_\delta t_{I,I+\delta}\left [ G_{I+\delta ,I}^{LDA(I)}(\omega)-\frac{\omega+\mu_{I+\delta}-\Sigma_{I+\delta ,I+\delta}^{LDA(I+\delta)}}
{\omega+\mu_{I+\delta}-\Sigma_{I+\delta ,I+\delta}(\omega)}G_{I+\delta ,I}^{LDA(I)}(\omega)\right ]\nonumber\\
&~&-\sum_\delta\frac{t_{I,I+\delta}t_{I+\delta, I}}{\omega+\mu_{I+\delta}-\Sigma_{I+\delta ,I+\delta}(\omega)} \left [ G_{I+\delta ,I+\delta}^{LDA(I+\delta)}(\omega)-\frac{\omega+\mu_I-\Sigma_{I,I}^{LDA(I)}(\omega)}{\omega+\mu_I-\Sigma_{I,I}(\omega)}G_{I,I}^{LDA(I)}(\omega)\right ]\Biggr \}
\label{eq: gga_dmft}
\end{eqnarray}
within the GGA-DMFT approach. For a spatially uniform system, we find $G$=$G^{LDA}$, same as before.
\end{widetext}

The LDA-DMFT algorithm is as follows: (1) determine all of the local chemical potentials $\{\mu_i=\mu-V_i\}$ for the different lattice sites; (2) start with a guess for the LDA self-energy on each lattice site (usually zero); (3) solve $|\Lambda|$ copies of the bulk problem for the LDA self-energy and LDA Green's function 
for each different value of $\mu_i$ via the standard homogeneous DMFT algorithm [(i) compute the local Green's function $G$ from the self-energy $\Sigma$ by summing over all momenta; (ii) extract the effective medium (or host Green's function) $G_0=1/(1/G+\Sigma)$; (iii) solve the impurity problem for the given effective medium to construct the new $G$; (iv) extract the new self-energy from the impurity Dyson's equation and the old effective medium and new Green's functions; (v) repeat (i-iv) until the equations reach a fixed point]; (4) use the new LDA self-energy on each lattice site to
initialize part (3) again and iterate until one reaches a fixed point; (5) compute the total density of particles and compare to the target density---if correct, then stop, if incorrect, update $\mu$ and repeat the process. Solving the LDA problem will produce the self-energies and the Green's functions for all
of the different lattice sites in the LDA approximation. These are then input into the GGA calculation, which involves the following steps;
(1) start with a guess for the self-energy given by the corresponding LDA solution---$\Sigma_{I,I}(\omega)=\Sigma^{LDA(I)}_{I,I}(\omega)$; (2) compute the new local Green's function $G$ at each site from
Eq.~(\ref{eq: gga_dmft}) and the fixed LDA results; (3) compute the effective medium for the 
impurity problem via $G_0=1/(1/G+\Sigma)$; (4) solve the impurity problem in the given effective medium;
(5) determine the new self-energy from the impurity Dyson's equation, the old effective medium and the new Green's function; (6) extract the self-energy via Dyson's equation; (7) iterate (2-6) until converged.
These solutions are the GGA-DMFT solutions to the problem. Note that this algorithm works for either real
or imaginary frequencies. Note further that the chemical potential should be adjusted to give the correct 
filling for the GGA. This implies that the LDA results employed within the GGA solution may be at the wrong chemical potential. This causes no problems with the formalism, but one should not compare the LDA solutions at this wrong chemical potential to the GGA solutions, rather one should compare the LDA and GGA solutions at the same total particle number.

\subsection{Application to the Falicov-Kimball model}

The spinless Falicov-Kimball model is given by the following Hamiltonian~\cite{falicov_kimball}
\begin{equation}
\mathcal{H}_{FK}=-\sum_{ij}t_{i,j}c^\dagger_ic^{\phantom\dagger}_j-\sum_i \mu_i c^\dagger_ic^{\phantom\dagger}_i+\sum_i E_i^fw_i+U\sum_i c^\dagger_ic^{\phantom\dagger}_i w_i
\label{eq: fk}
\end{equation}
where $\mu_i=\mu-V_i$ is the local chemical potential (and $V_i$ is the harmonic trap), $E_i^f=E^f+V^f_i$ is the local site energy for the static particles with $V^f_i$ the harmonic trap potential for the heavy particles, $U$ is the on-site interaction between heavy and light particles, $c_i^\dagger$ 
($c_i^{\phantom\dagger}$) are the fermionic creation and annihilation operators for the mobile electrons and $w_i=0$ or 1 is the heavy particle number operator. This Hamiltonian can be thought of as the
Hubbard model Hamiltonian with a spin-dependent hopping and the hopping for the down particles set to zero.

\begin{figure*}
\centerline{\includegraphics [width=2.3in, angle=0, clip=on]  {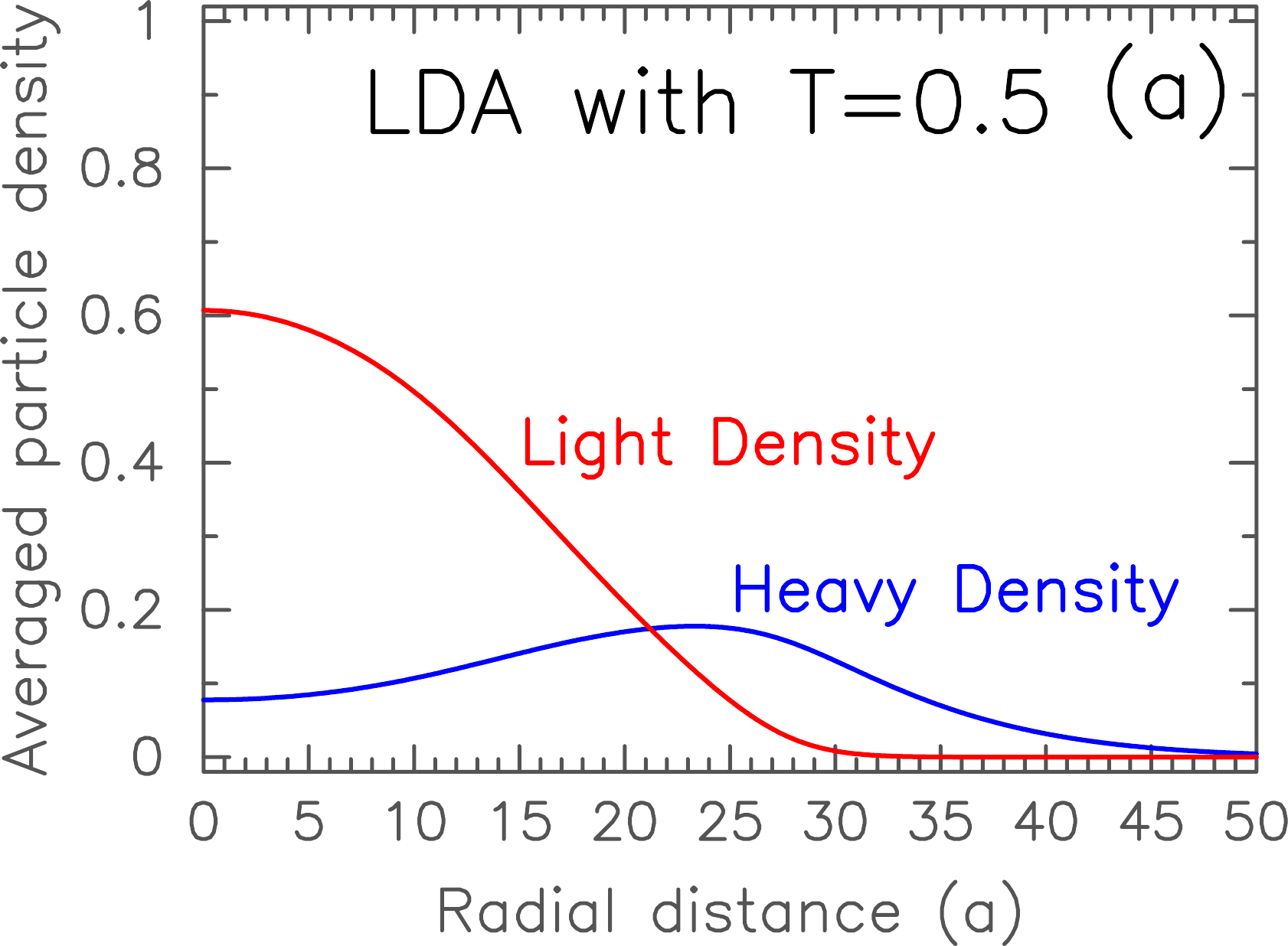}
\includegraphics [width=2.3in, angle=0, clip=on]  {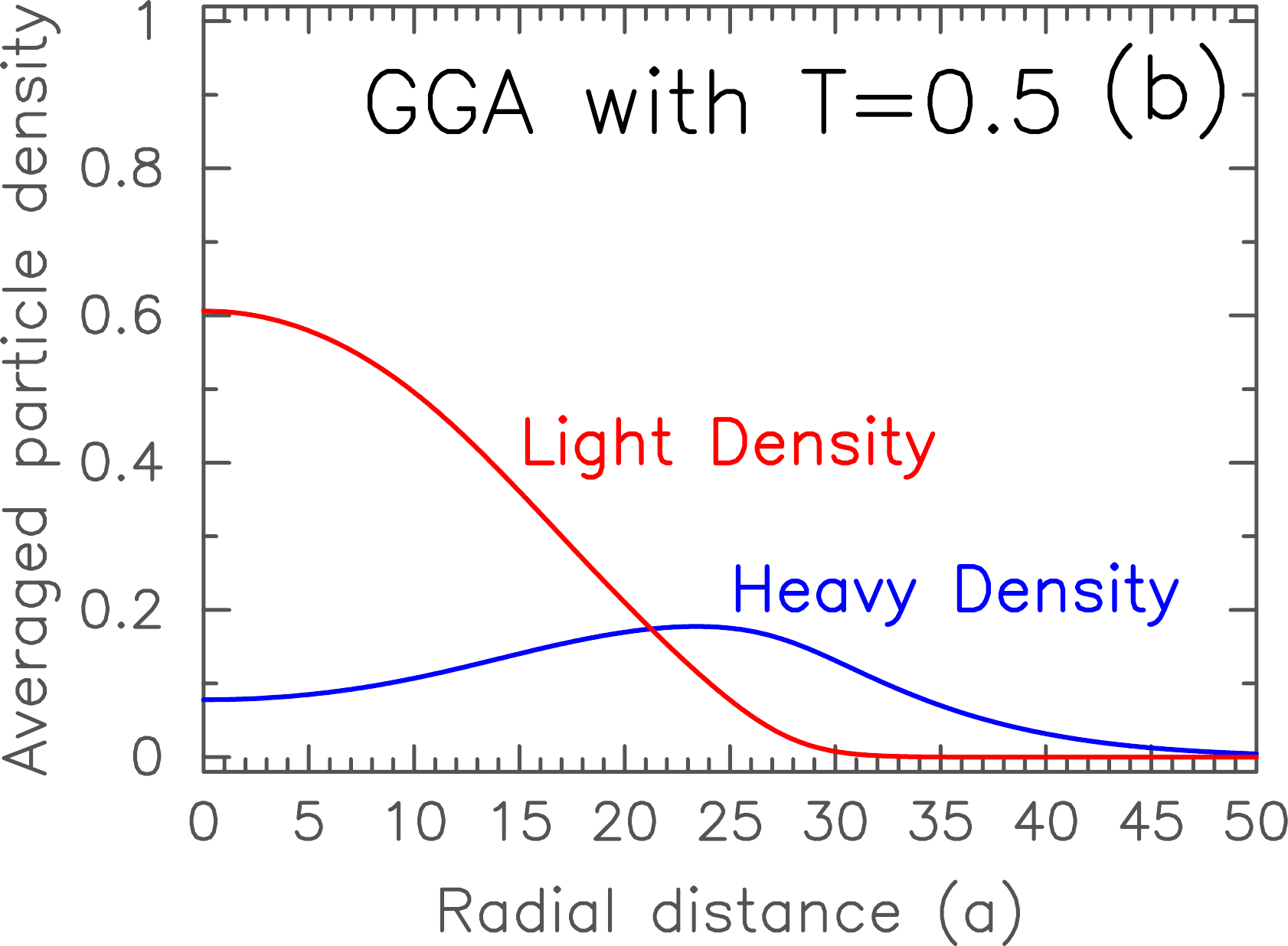}
\includegraphics [width=2.3in, angle=0, clip=on]  {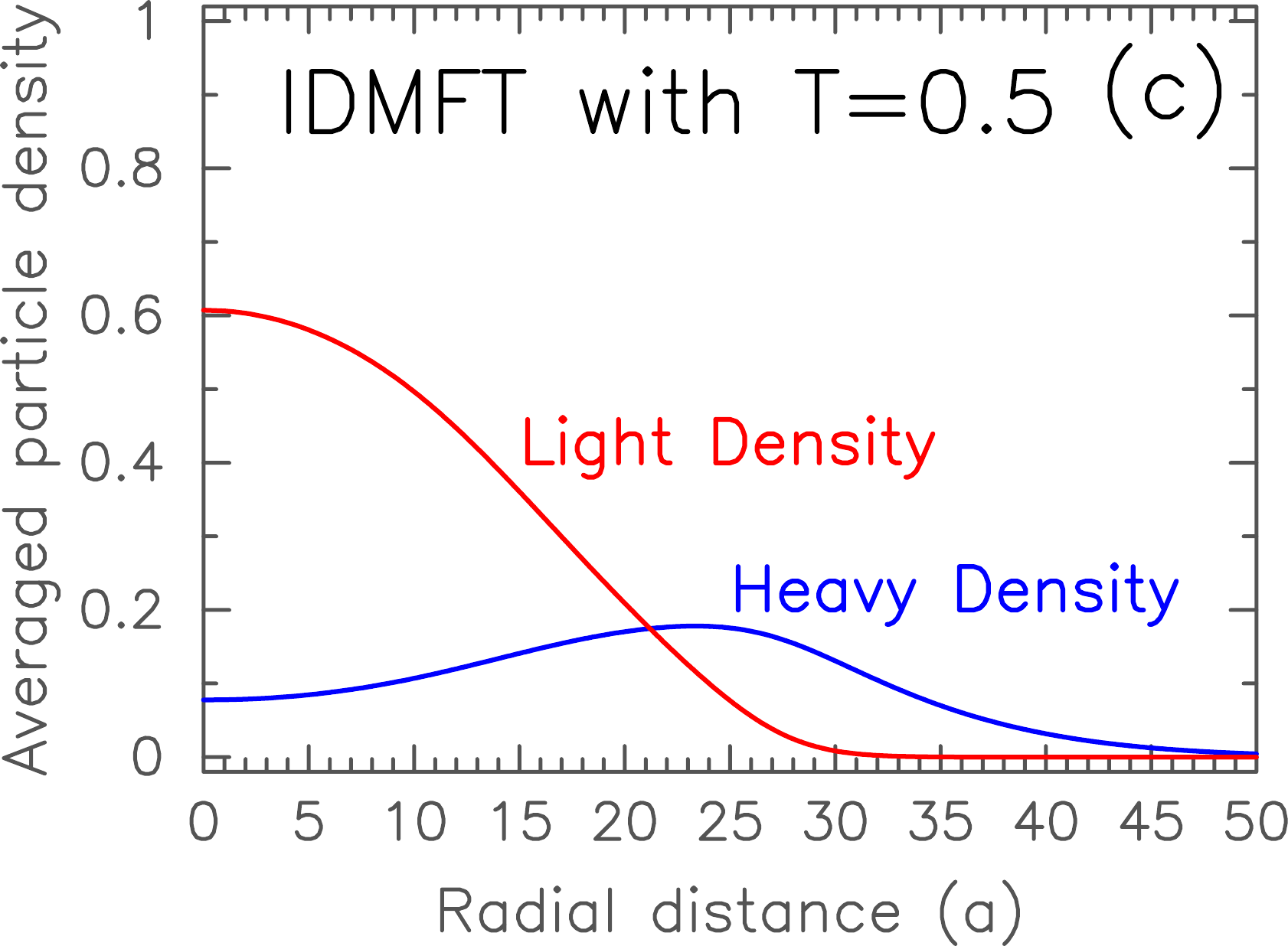}
}
\centerline{\includegraphics [width=2.3in, angle=0, clip=on]  {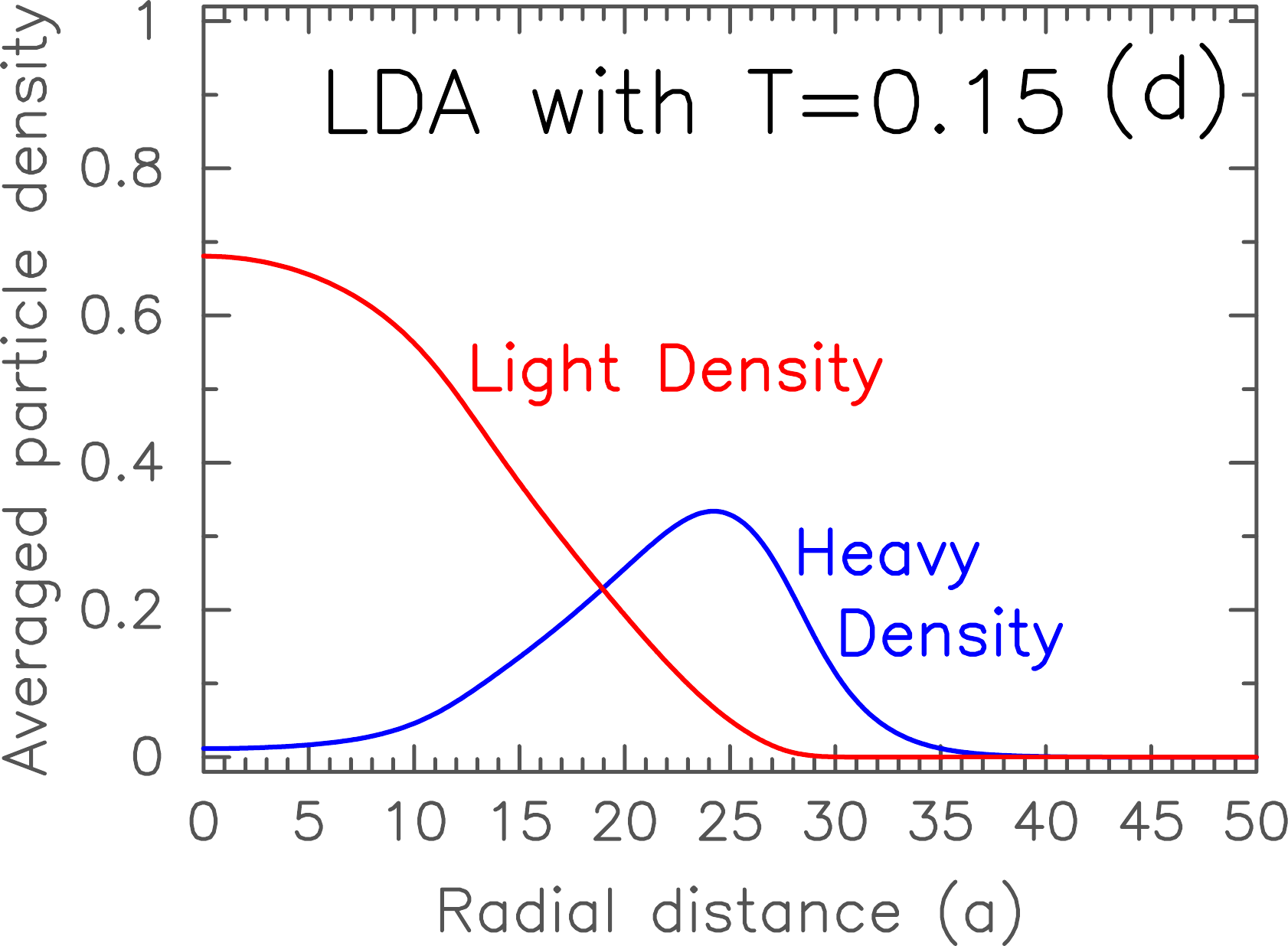}
\includegraphics [width=2.3in, angle=0, clip=on]  {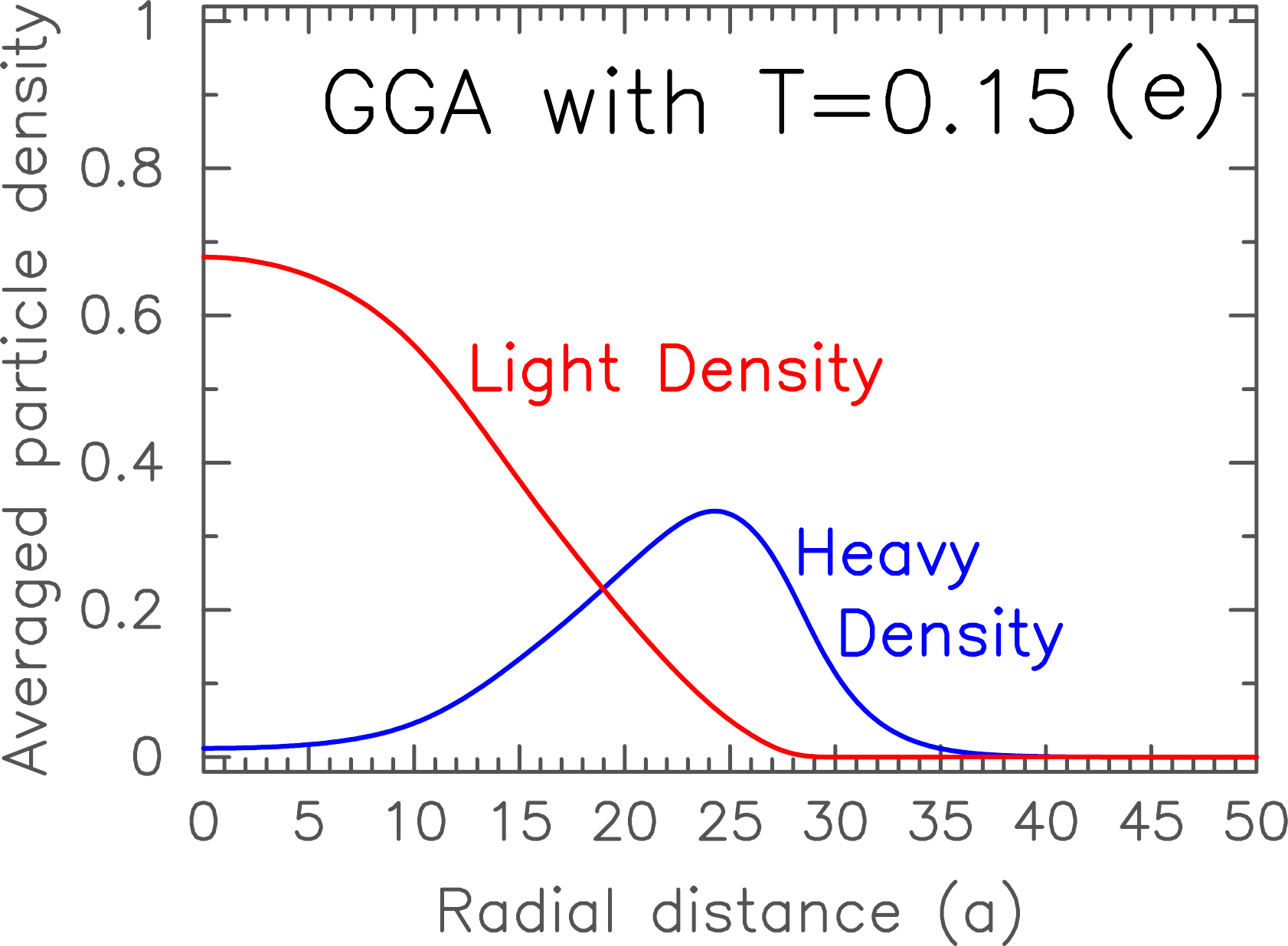}
\includegraphics [width=2.3in, angle=0, clip=on]  {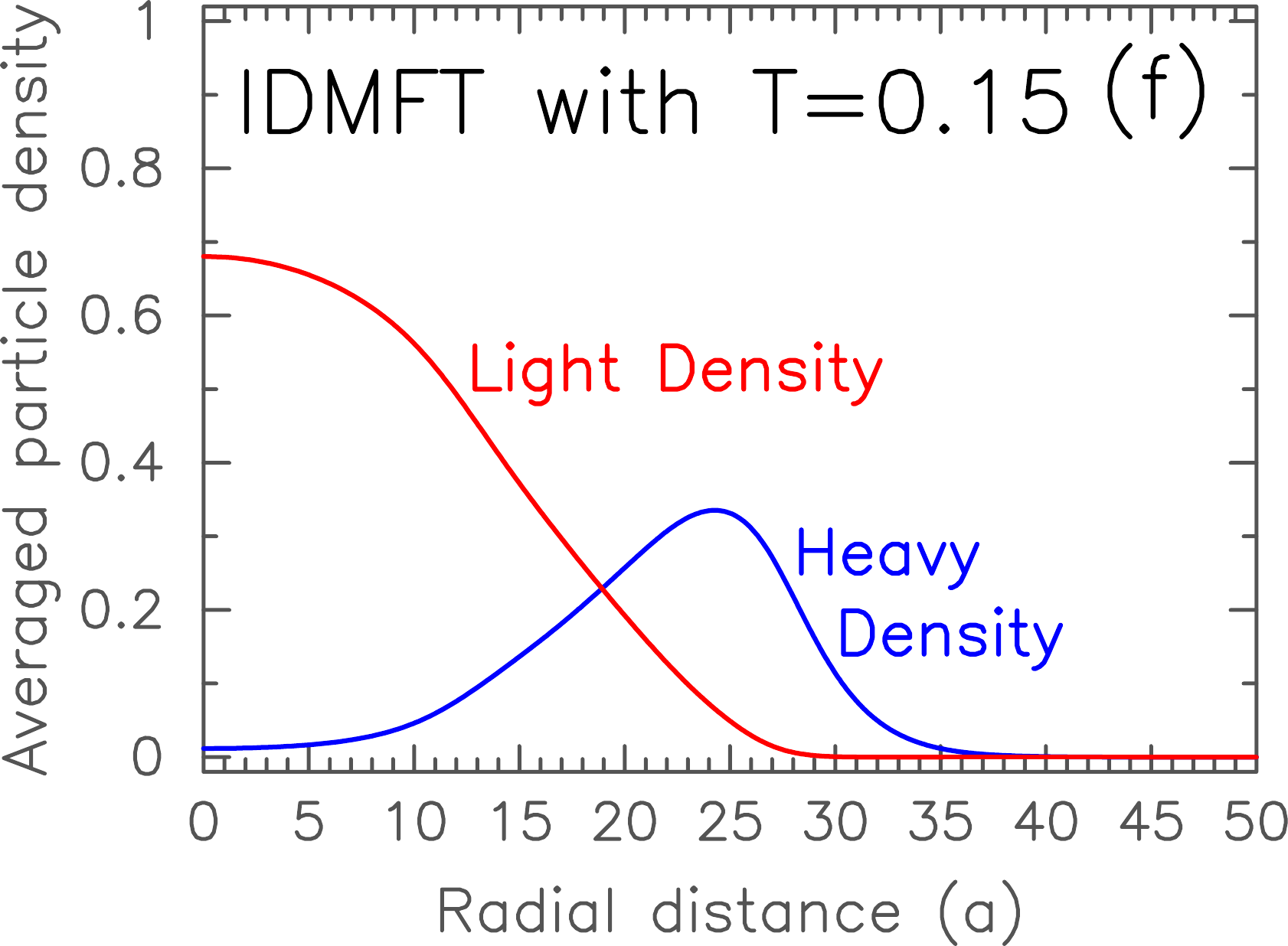}
}
\caption[]{
(Color on-line.) (Upper row) Radial density for T=0.5. Blue is for heavy particles, red is for light particles. From left to right, we have the (a) LDA, (b) GGA, and (c) IDMFT results.
(Lower row) Radial density for T=0.15. Blue is for heavy particles, red is for light particles. From left to right, we have the (d) LDA, (e)  GGA, and (f) IDMFT results.
\label{fig: highT}
}
\end{figure*}

The dynamical mean-field theory  for the Falicov-Kimball model is exactly solvable~\cite{brandt_mielsch,freericks_review}. We usually solve the problem on the imaginary axis to determine the chemical potentials, and the fillings of the different particles on the different sites.  Many other quantities can be determined from these solutions, like the total energy, kinetic energy, and so on; but some observables, like the entropy, require results on the real frequency axis, and so one would then repeat the above procedure, this time for the real frequency axis, to calculate those 
observables. In that case, the chemical potentials would be set from the imaginary axis calculations, and the real-axis calculations would be done primarily to find the spectral functions, and quantities that can be derived from them.

One technical element for the Falicov-Kimball model on the imaginary axis, is that it is useful to sum
the tails of series using exact sum rules for summations of Green's functions. These sum rules can be found from the moments of the Green's functions. We use this technique to improve the accuracy of our calculations when we determine the filling on each lattice site. For all of the properties reported here with the Falicov-Kimball model, we focused entirely on the imaginary-axis results---no real-axis calculations were needed.

Because we will be comparing these LDA-DMFT and GGA-DMFT results for the Falicov-Kimball model to a complete inhomogeneous DMFT calculation, we cannot 
make the system size too large.  We work primarily with about 10,000 lattice sites and total particle number equal to 1350 (625 for each species). This system is slightly smaller than a typical experimental system, but only by a factor of four or so. Note also that earlier work already showed that the IDMFT
solution agrees well with the Monte Carlo solution~\cite{maska}.

\begin{figure*}
\centerline{\includegraphics [width=2.3in, angle=0, clip=on]  {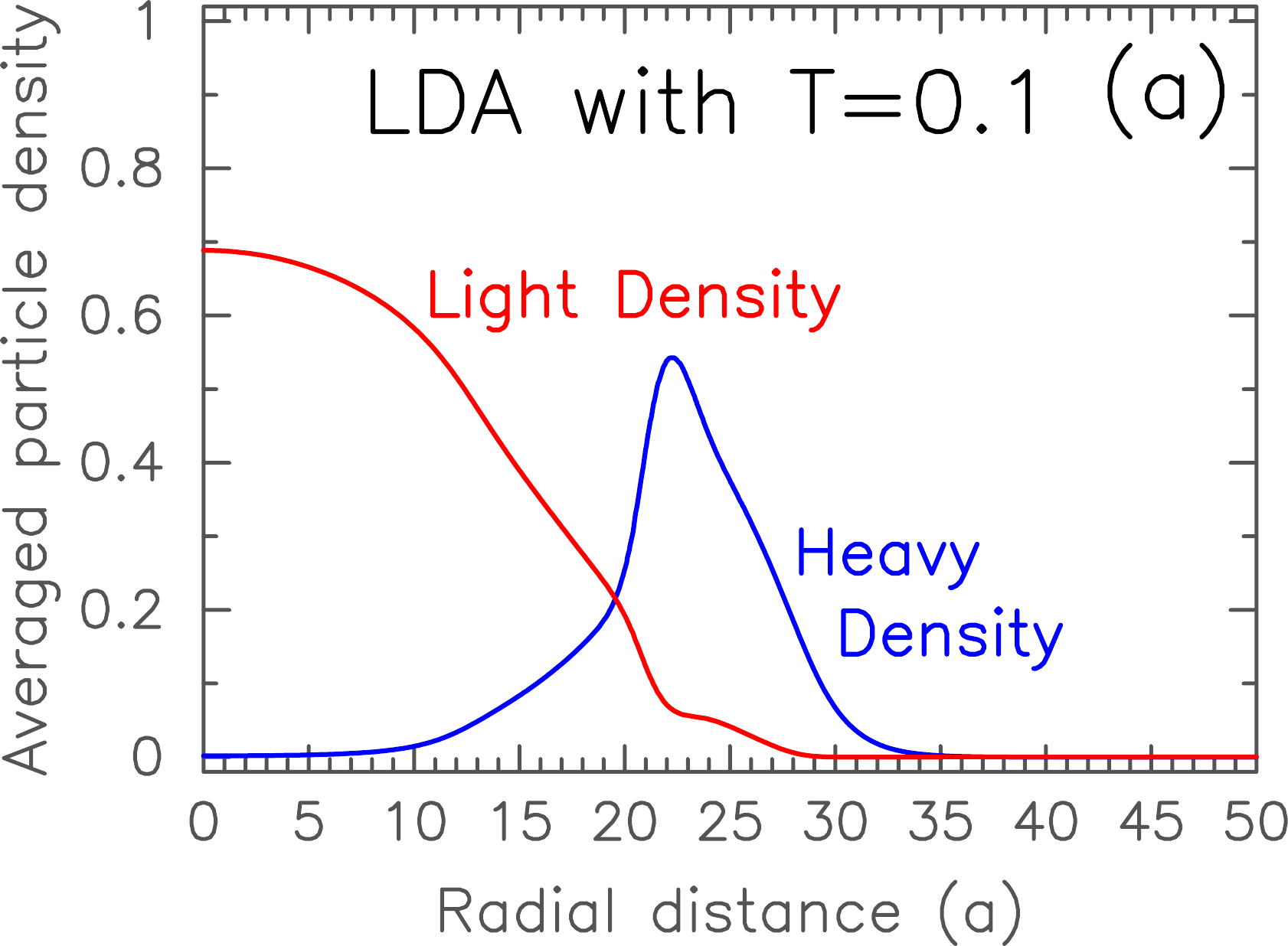}
\includegraphics [width=2.3in, angle=0, clip=on]  {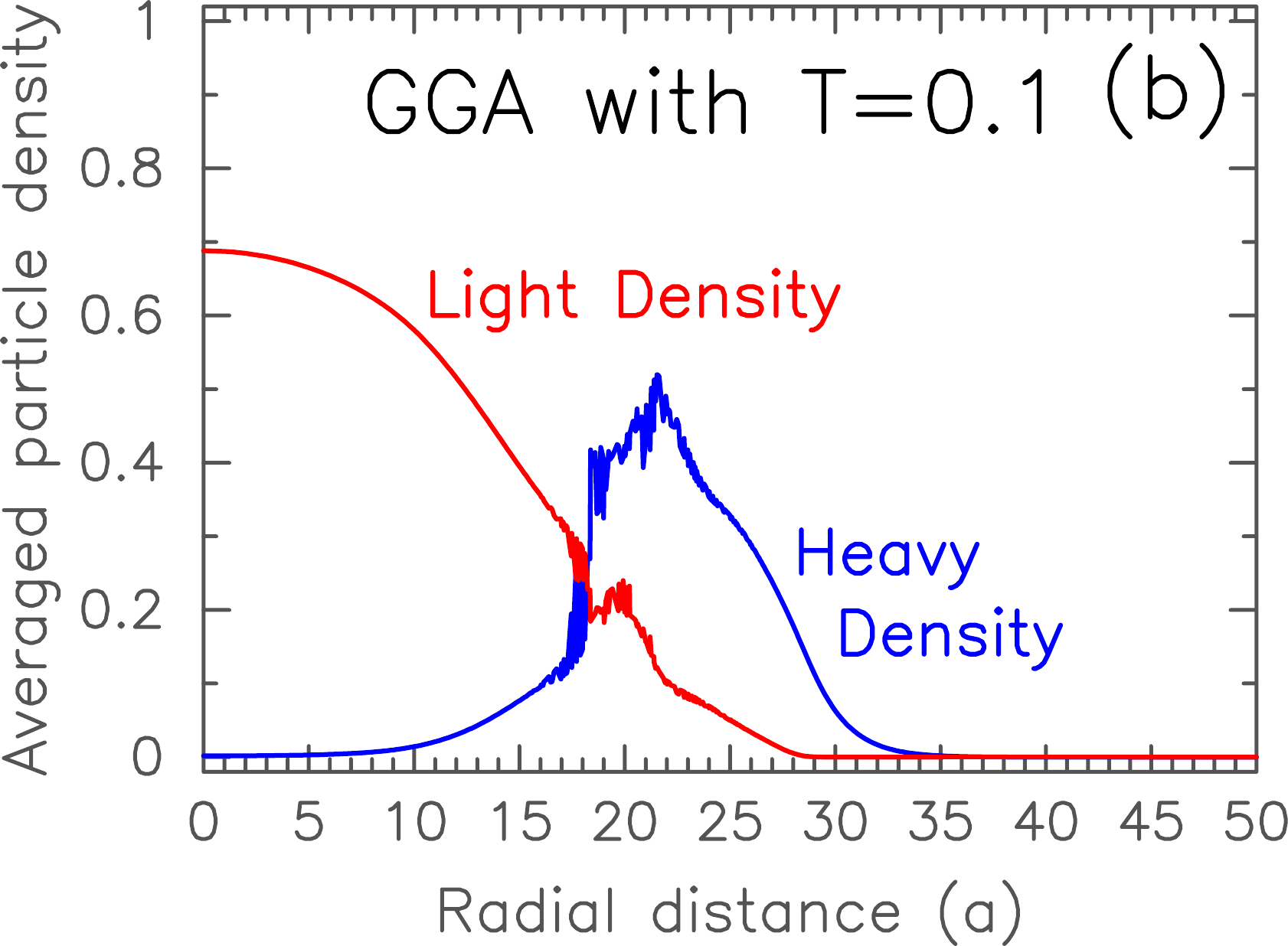}
\includegraphics [width=2.3in, angle=0, clip=on]  {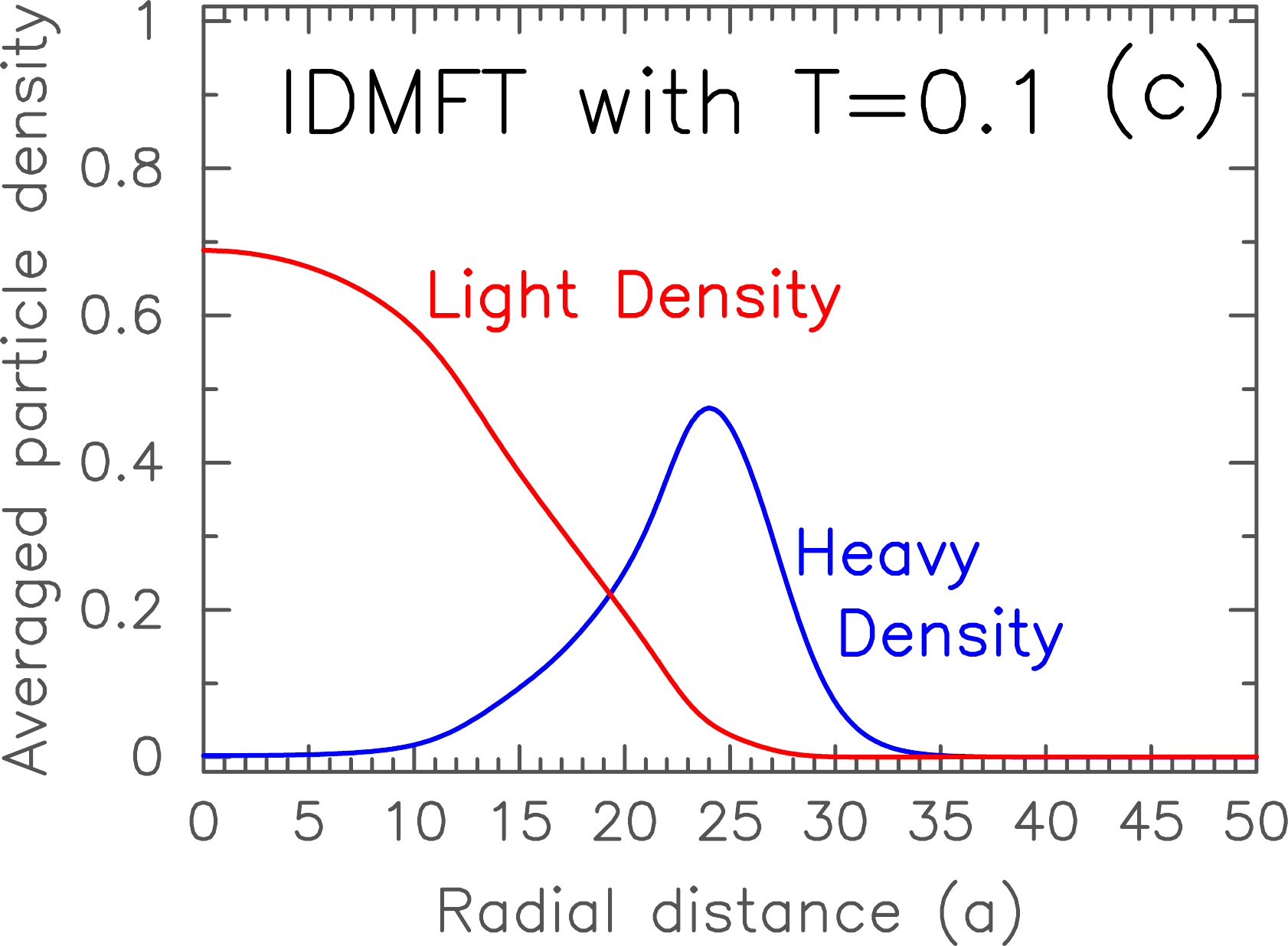}
}
\centerline{\includegraphics [width=2.3in, angle=0, clip=on]  {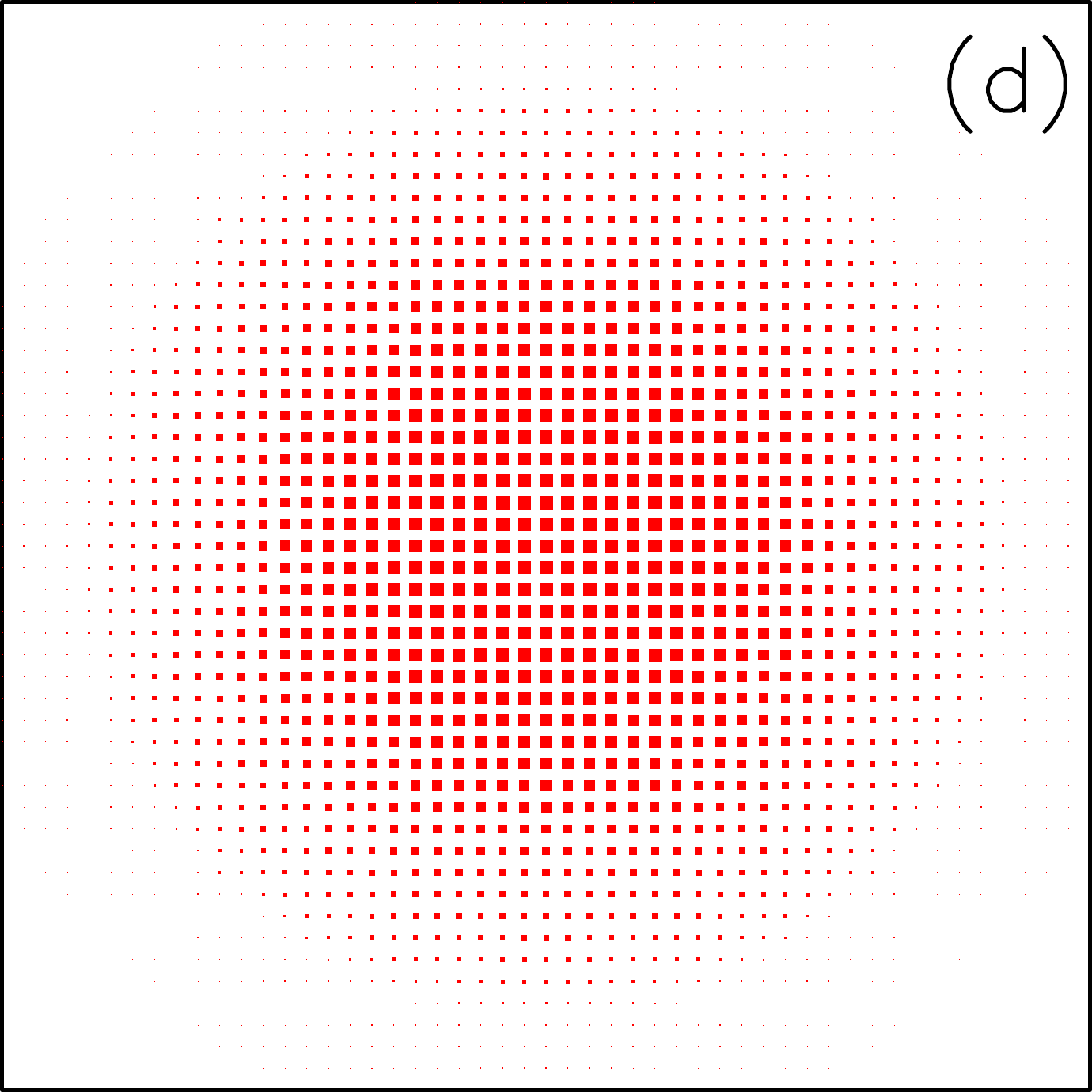}
\includegraphics [width=2.3in, angle=0, clip=on]  {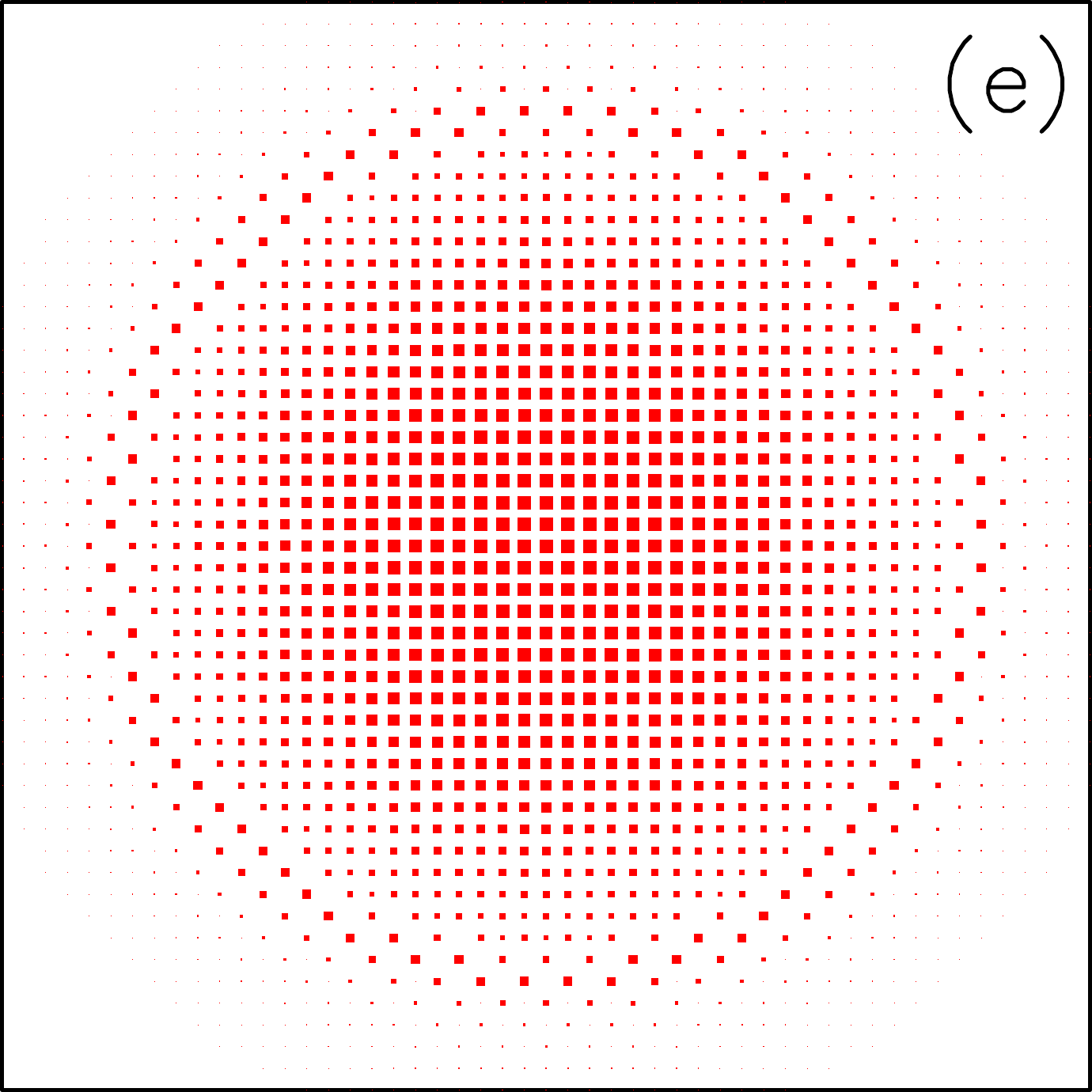}
\includegraphics [width=2.3in, angle=0, clip=on]  {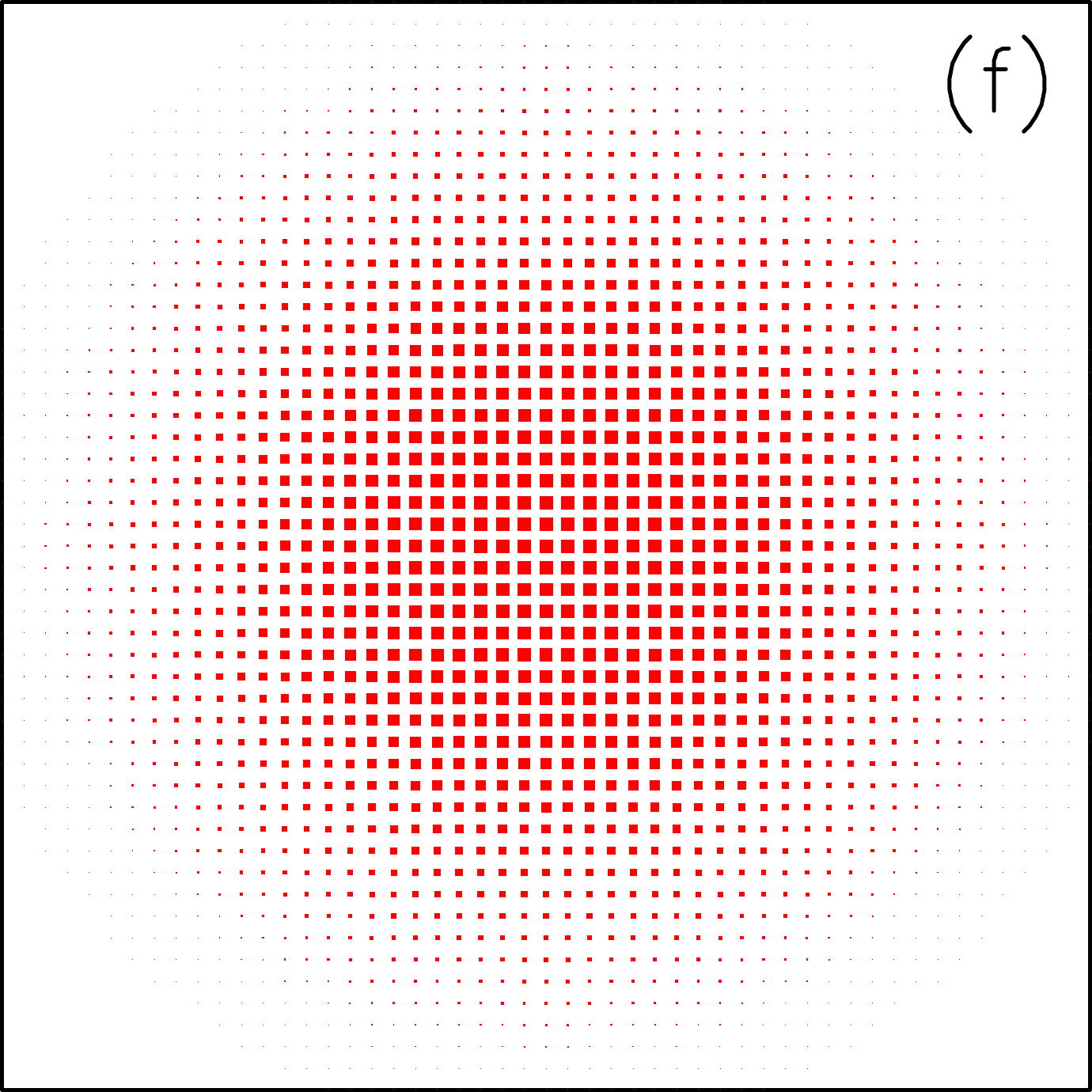}
}
\centerline{\includegraphics [width=2.3in, angle=0, clip=on]  {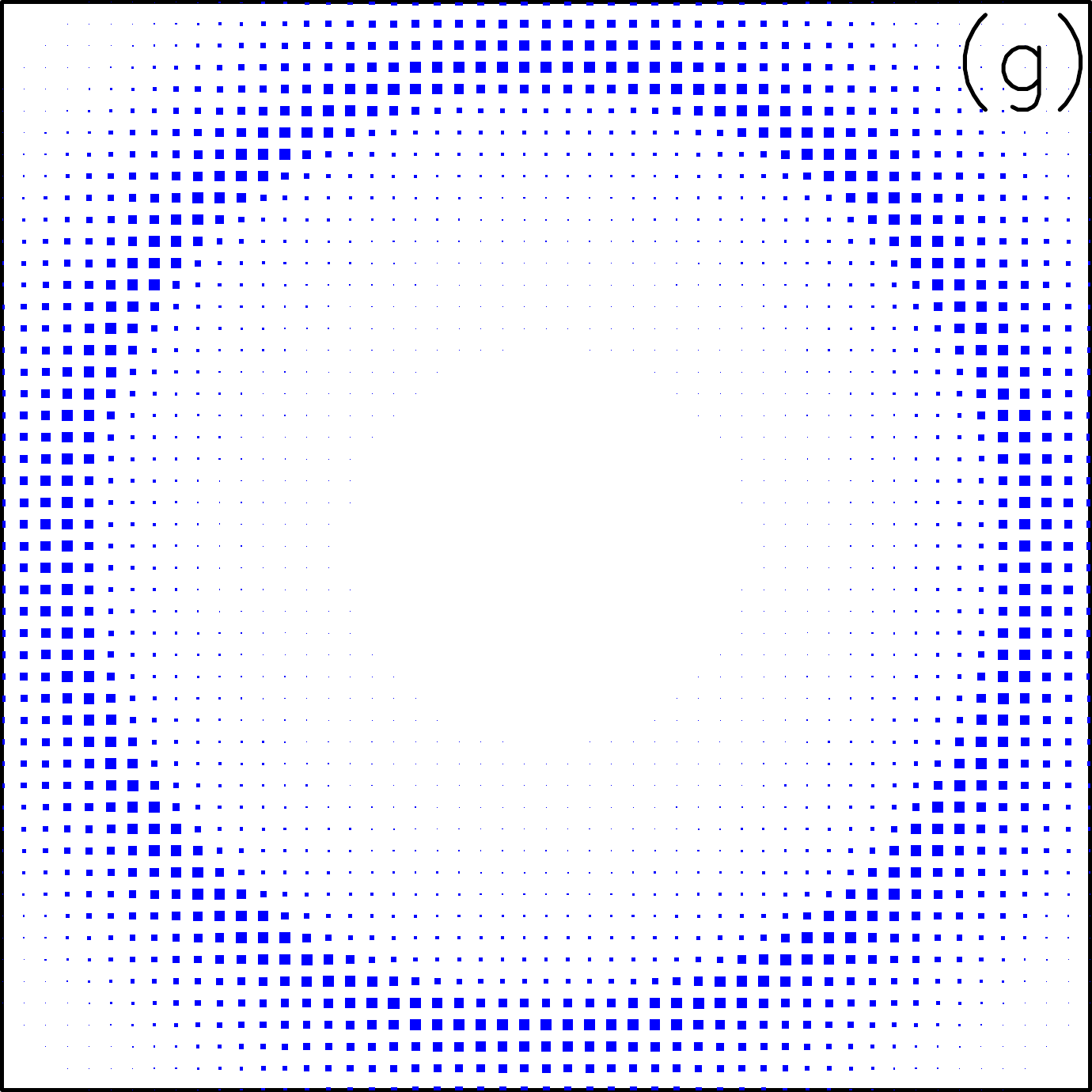}
\includegraphics [width=2.3in, angle=0, clip=on]  {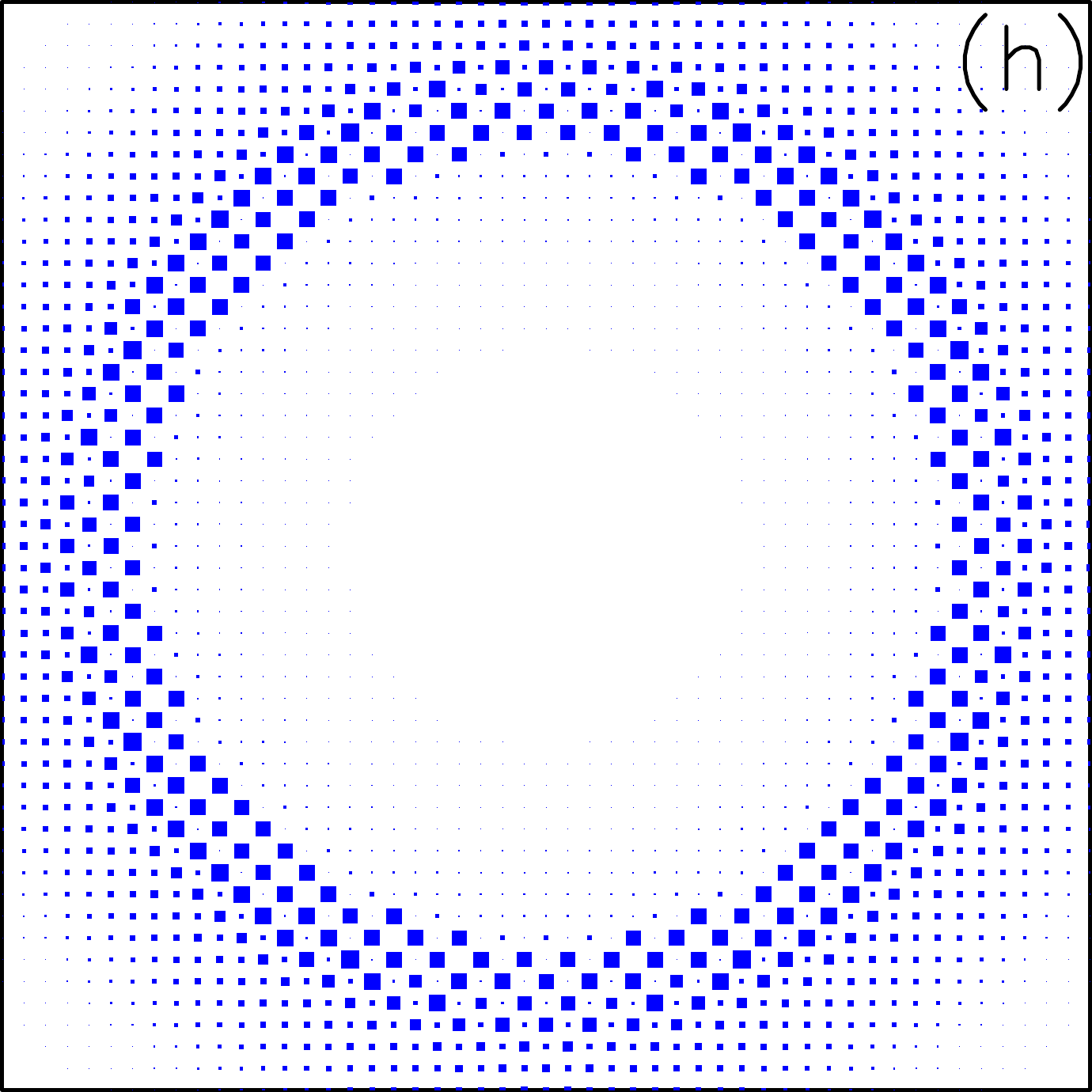}
\includegraphics [width=2.3in, angle=0, clip=on]  {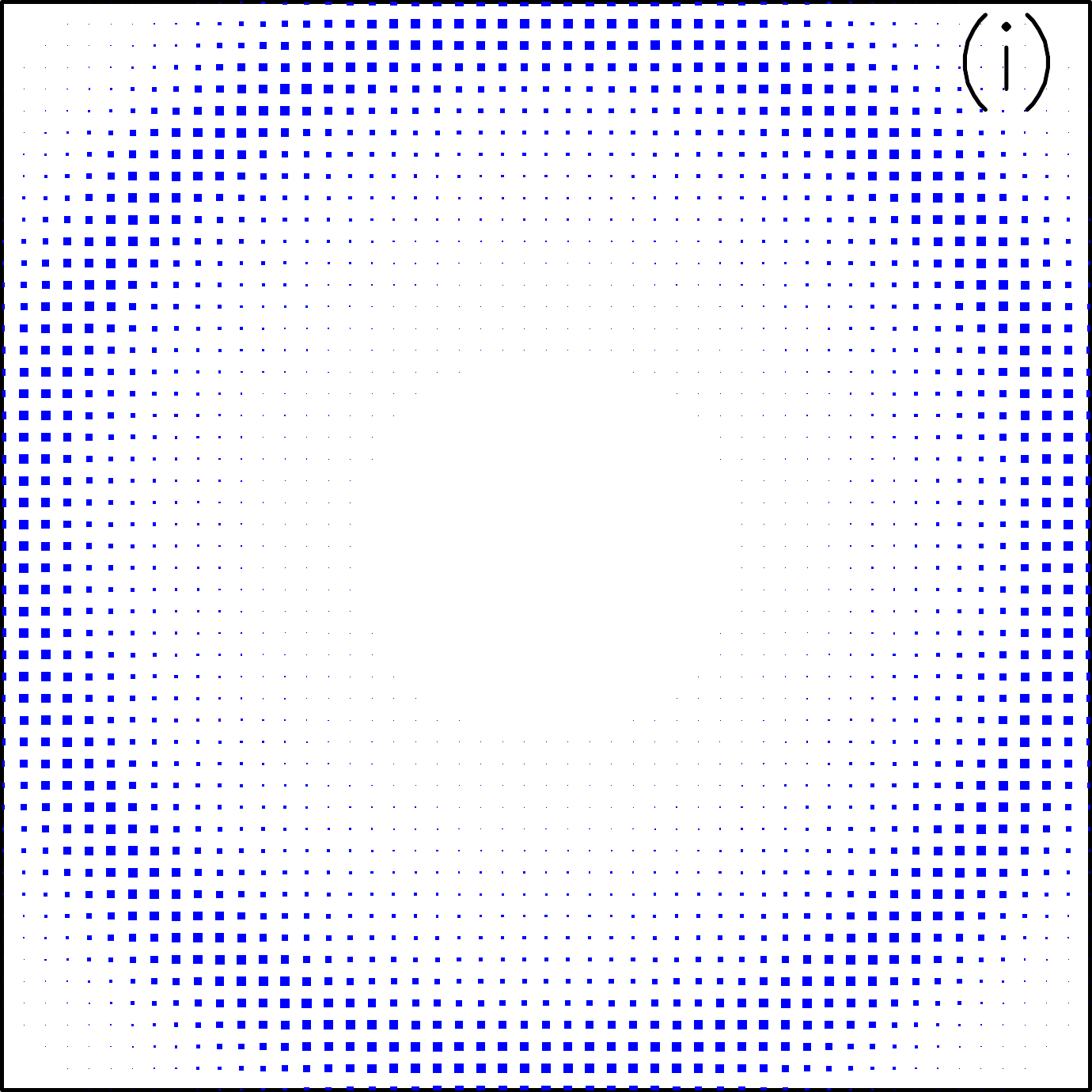}
}
\caption[]{
(Color on-line.) (Upper row) Radial density for T=0.1. Blue is for heavy particles, red is for light particles. From left to right, we have the (a) LDA, (b) GGA, and (c) IDMFT results.
(Middle row) Light particle density for T=0.1 (red). The size of the symbol is proportional to the density of the light particles at that site. From left to right, we have the (d) LDA, (e) GGA, and (f) IDMFT results.
(Lower row) Heavy particle density for T=0.1 (blue). The size of the symbol is proportional to the density of the heavy particles at that site. From left to right, we have the (g) LDA, (h) GGA, and (i) IDMFT results.
\label{fig: t=0.1}
}
\end{figure*}

\subsection{Application to the Fermi Hubbard model}

The Hubbard model Hamiltonian is~\cite{hubbard}
\begin{equation}
\mathcal{H}_{H}=-\sum_{ij\sigma}t_{i,j;\sigma}c^\dagger_{i\sigma}c^{\phantom\dagger}_{j\sigma}-\sum_i \mu_i c^\dagger_{i\sigma}c^{\phantom\dagger}_{i\sigma}+U\sum_i c^\dagger_{i\uparrow}c^{\phantom\dagger}_{i\uparrow}c^\dagger_{i\downarrow}c^{\phantom\dagger}_{i\downarrow}
\end{equation}
where we have mobile up and down spin particles now.

We employ two different techniques to solve this problem. One is a strong-coupling perturbation theory
approach (through second-order in the hopping divided by the interaction), which has been used earlier to determine the inhomogeneous distribution of the density of the particles and the entropy per particle of the system~\cite{scarola}. We do not repeat the derivation of those formulas here.

We also use a continuous-time quantum Monte Carlo 
approach with a weak-coupling implementation. This approach uses stochastic sampling to sum a series of Feynman diagrams in imaginary-time which then solves the problem~\cite{ctqmc1,ctqmc2}. The algorithm sums over a random collection of diagrams at different orders, and adjusts the order of the diagram based on importance sampling via a Metropolis strategy. In the implementation that we employ, the approach is most accurate at high-temperature and small interaction strengths. As both the temperature is reduced and the interaction is increased, the average order of the calculation increases, and the integration range in imaginary-time also increases (which requires more sampling). This technique is currently believed to be
the most accurate state-of-the-art approach for determining properties that can be calculated at finite-temperatures using Green's functions evaluated along the imaginary-time axis. These include the particle-density at each lattice site, the double occupancy, the entropy-per-particle, and the order parameter if the system goes into an ordered antiferromagnetic phase.
Details for how this algorithm is implemented can be found elsewhere.~\cite{ctqmc3}

\begin{figure*}
\centerline{\includegraphics [width=2.3in, angle=0, clip=on]  {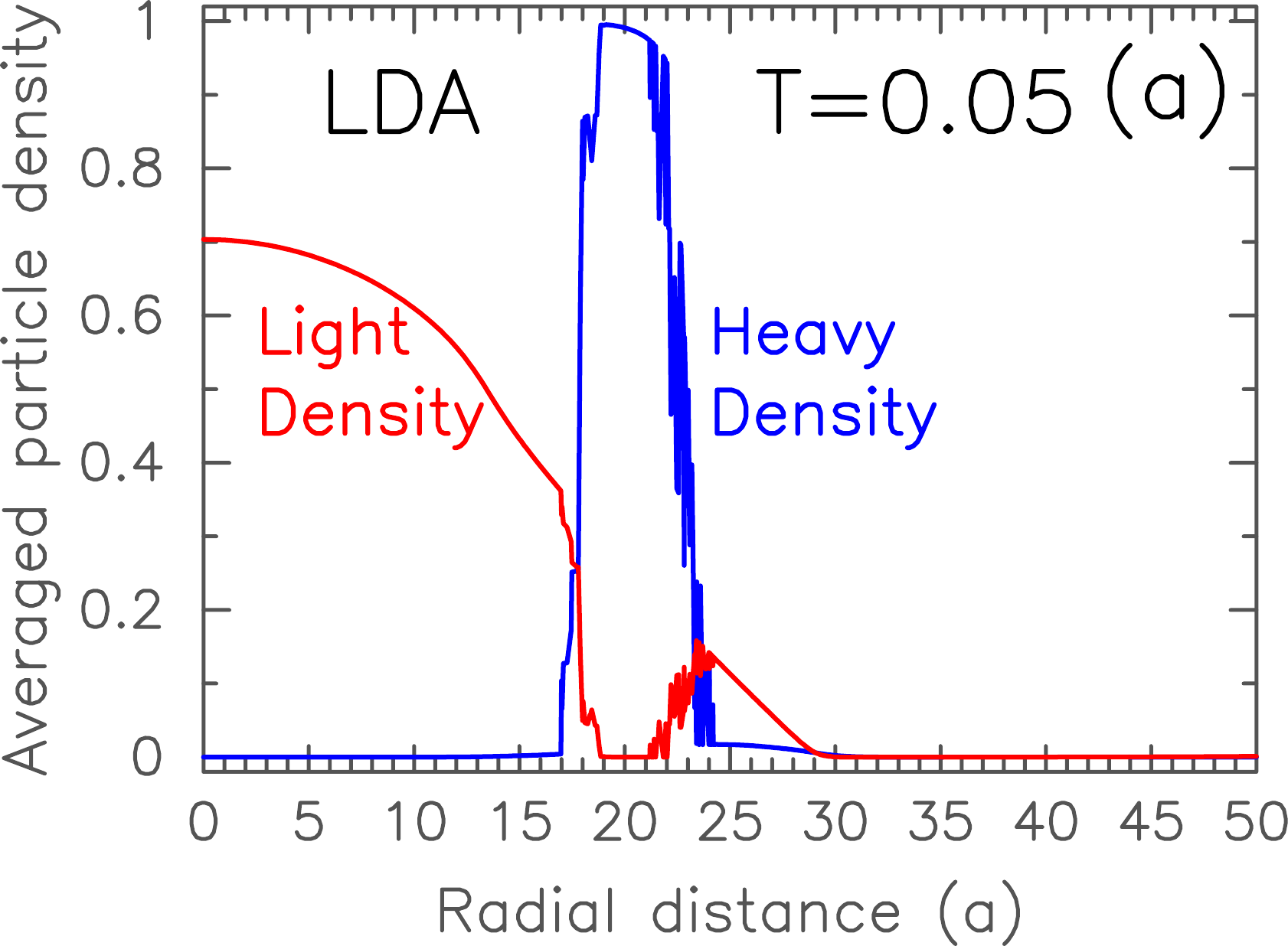}
\includegraphics [width=2.3in, angle=0, clip=on]  {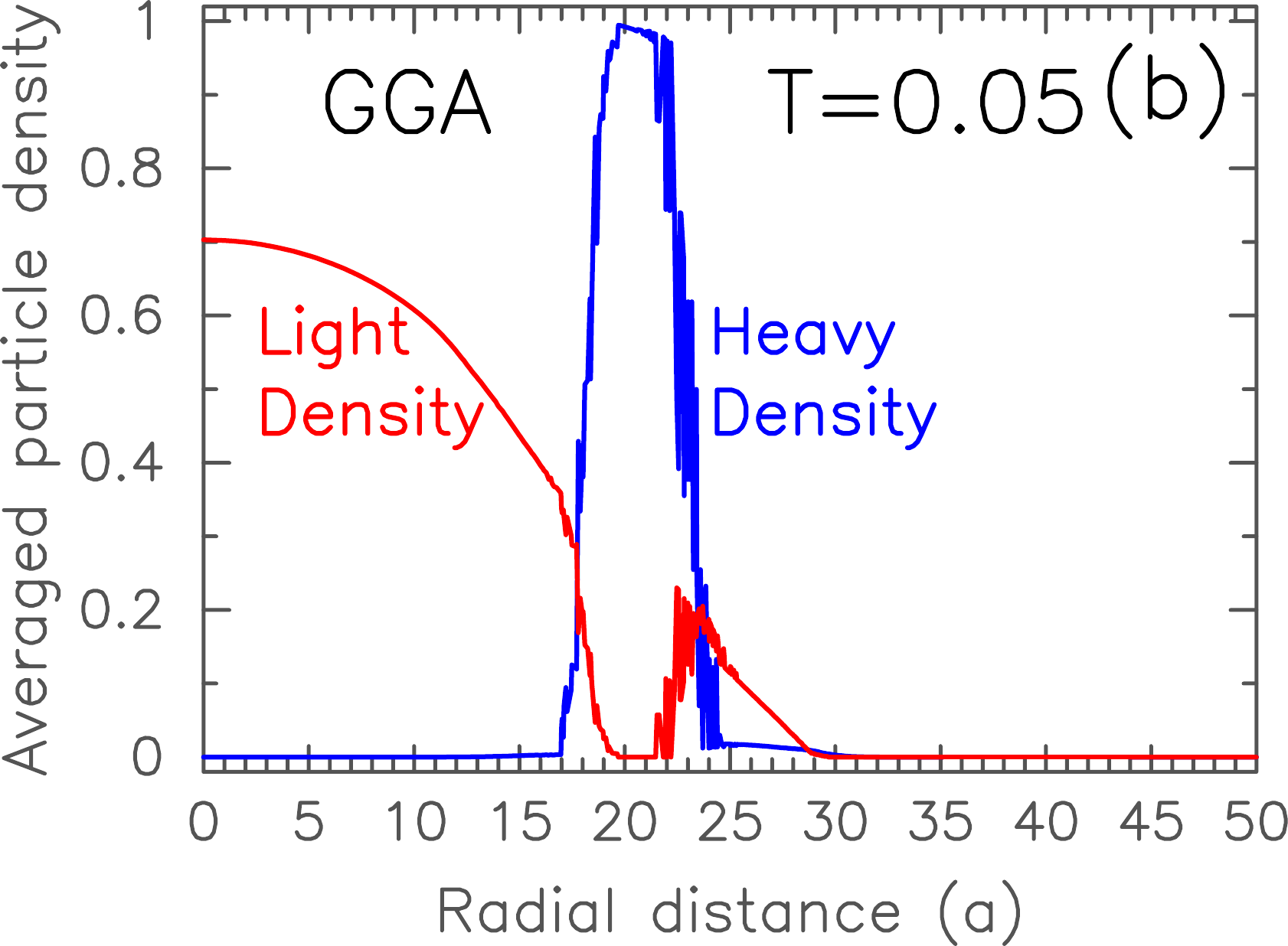}
\includegraphics [width=2.3in, angle=0, clip=on]  {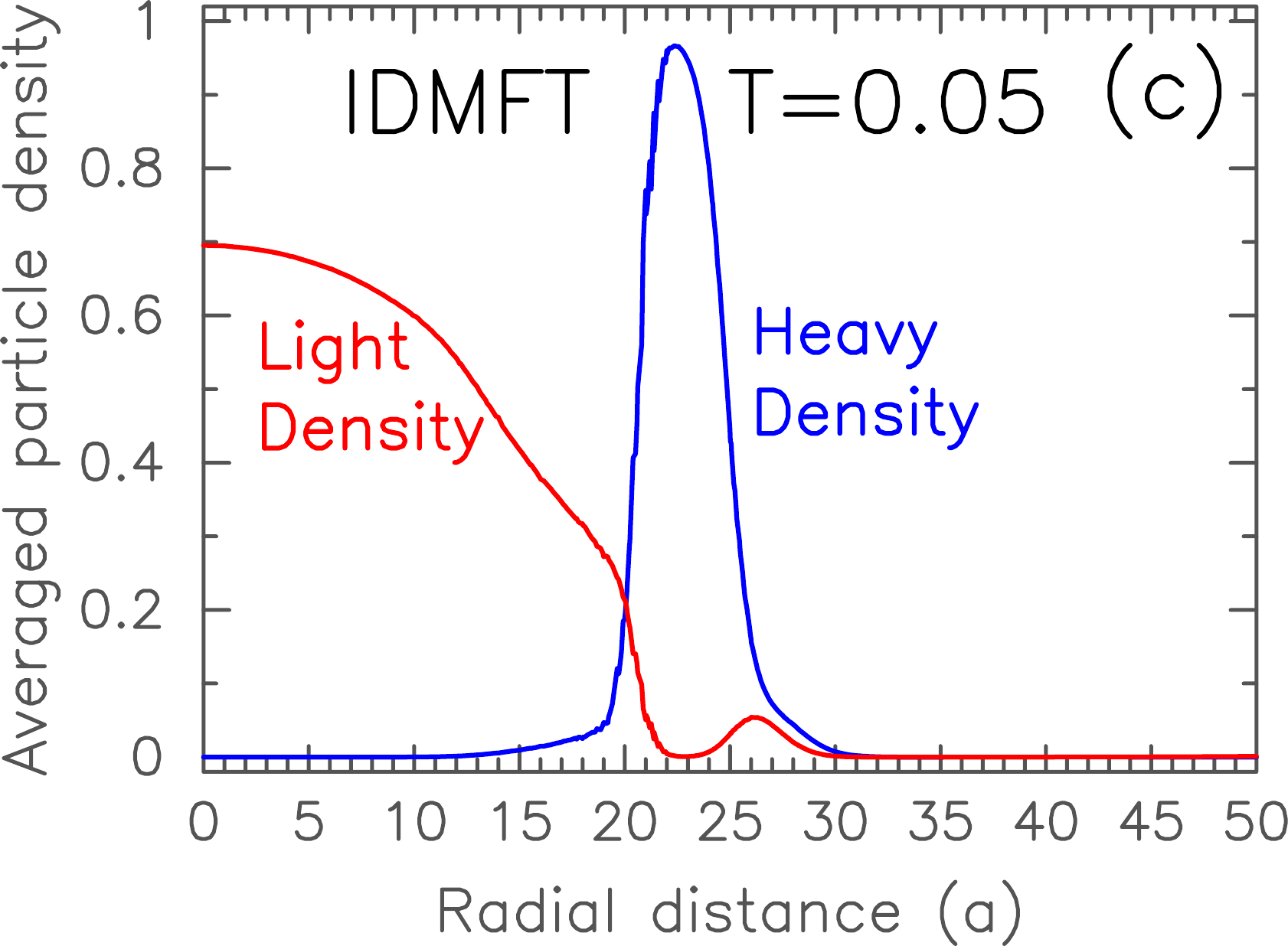}
}
\centerline{\includegraphics [width=2.3in, angle=0, clip=on]  {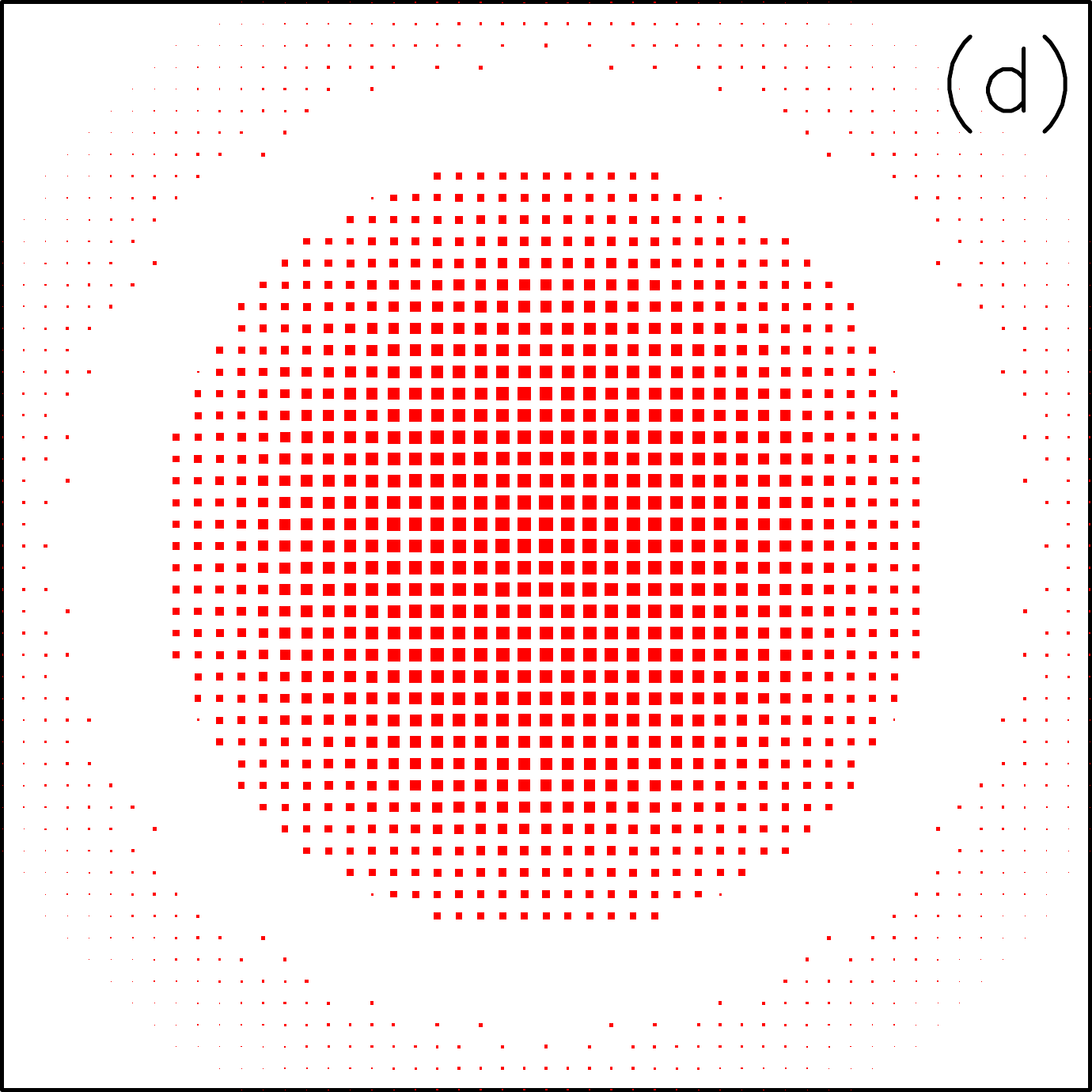}
\includegraphics [width=2.3in, angle=0, clip=on]  {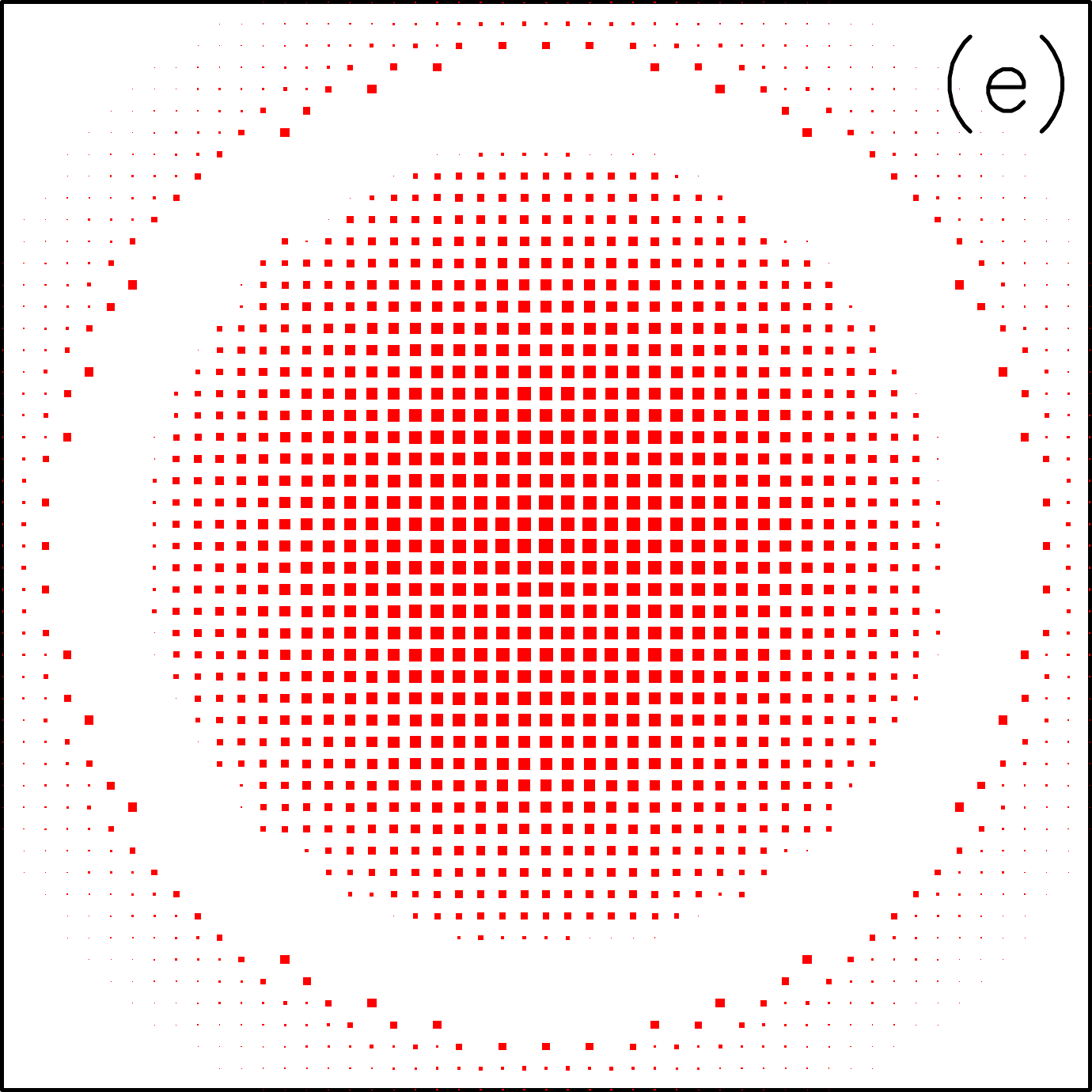}
\includegraphics [width=2.3in, angle=0, clip=on]  {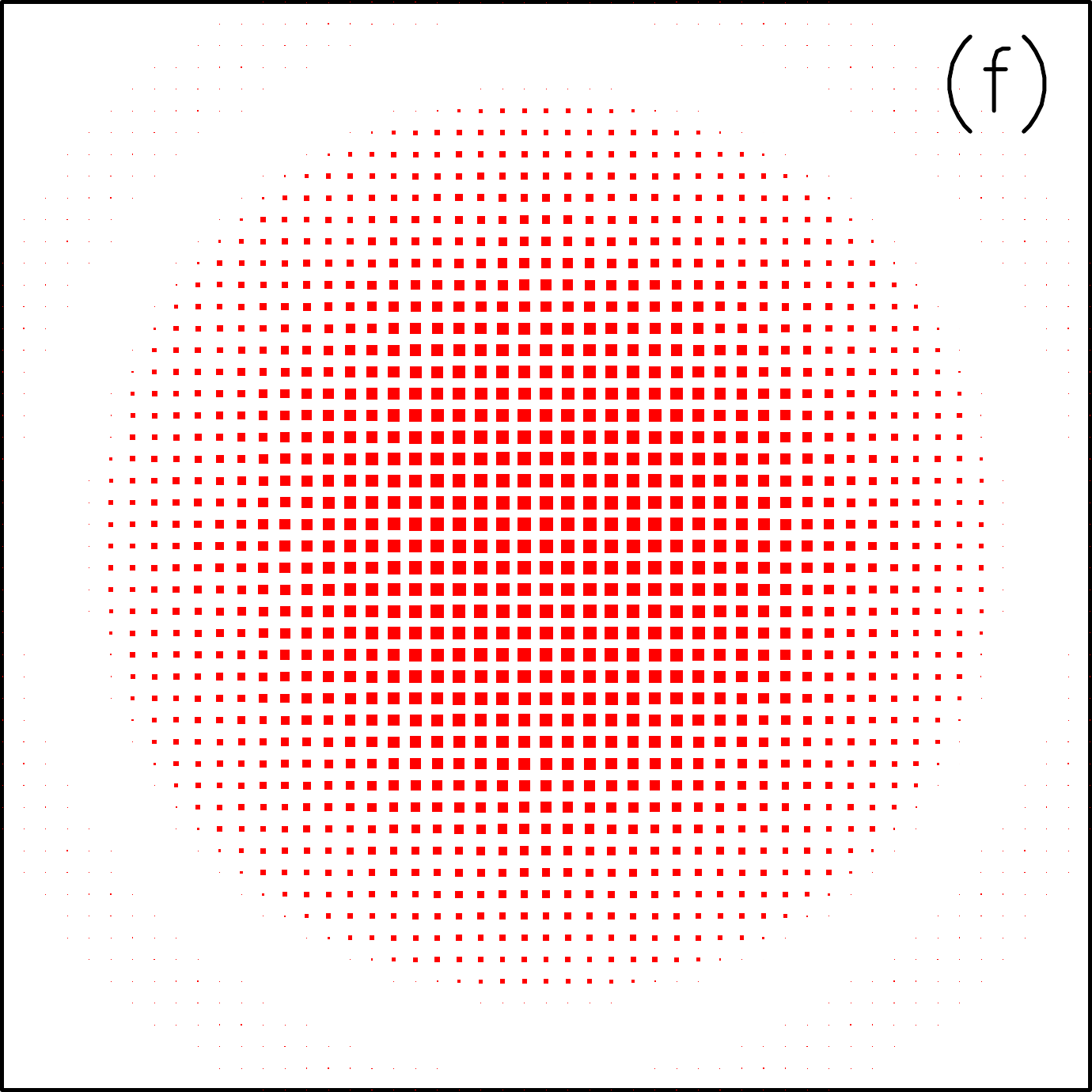}
}
\centerline{\includegraphics [width=2.3in, angle=0, clip=on]  {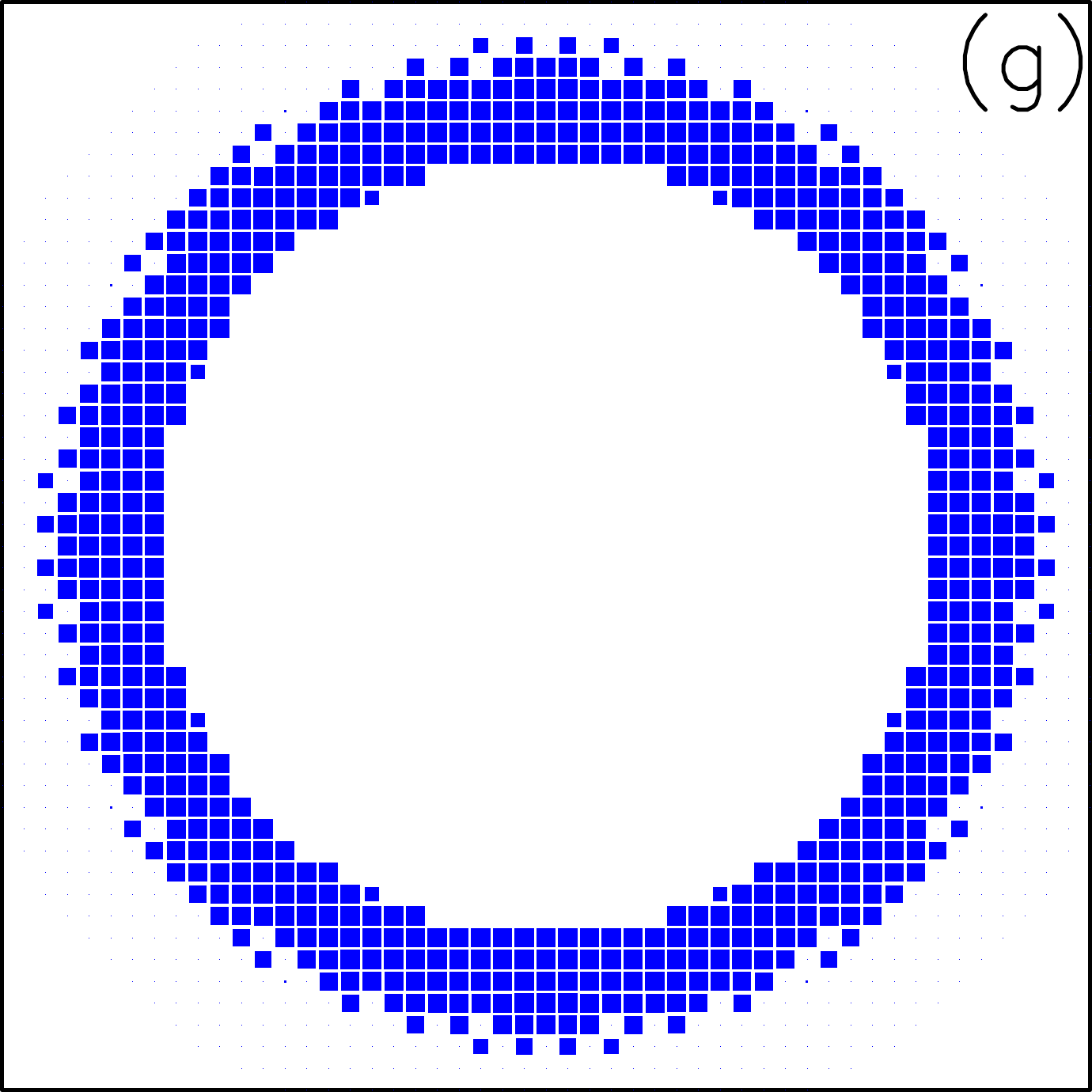}
\includegraphics [width=2.3in, angle=0, clip=on]  {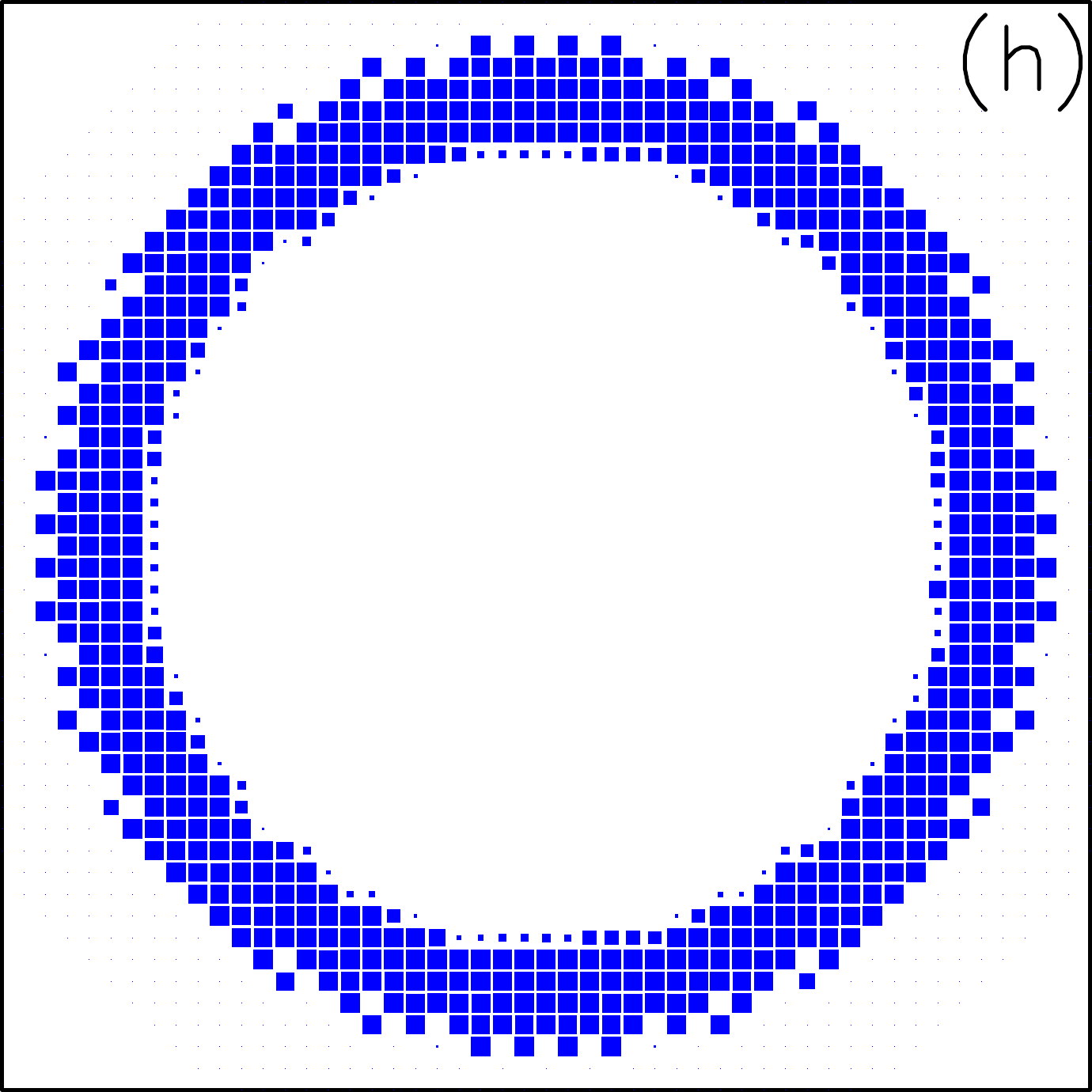}
\includegraphics [width=2.3in, angle=0, clip=on]  {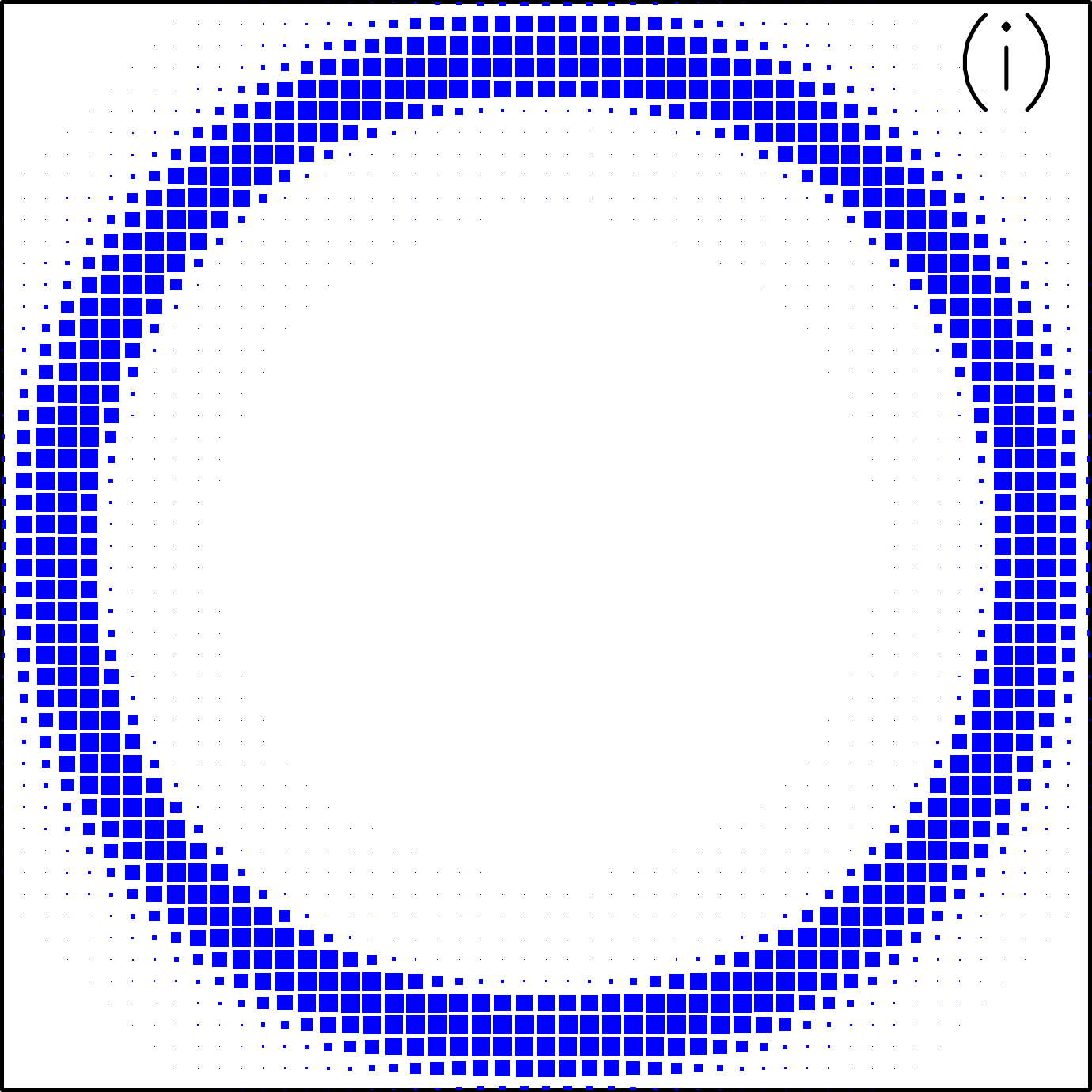}
}
\caption[]{
(Color on-line.) (Upper row) Radial density for T=0.05. Blue is for heavy particles, red is for light particles. From left to right, we have the (a) LDA, (b) GGA, and (c) IDMFT results.
(Middle row) Light particle density for T=0.05 (red). The size of the symbol is proportional to the density of the light particles at that site. From left to right, we have the (d)  LDA, (e) GGA, and (f) IDMFT results.
(Lower row) Heavy particle density for T=0.05 (blue). The size of the symbol is proportional to the density of the heavy particles at that site. From left to right, we have the (g) LDA, (h) GGA, and (i) IDMFT results.
\label{fig: t=0.05}
}
\end{figure*}

Here, we focus on the density distribution and the calculation of the total energy of the system.  Starting from high temperature, we can then integrate to find the entropy via the relation~\cite{werner}
\begin{eqnarray}
S(\beta)&=&S(0)+\int_0^\beta \bar\beta \frac{dE(\bar\beta)}{d\bar\beta} d\bar\beta\nonumber\\
&\approx& S(\beta^\prime)+\frac{1}{2}[E(\beta)-E(\beta^\prime)](\beta+\beta^\prime),
\end{eqnarray}
for $\beta-\beta^\prime$ small, with $S(0)$ the infinite-temperature entropy. This result follows from the thermodynamic relation $\partial_T E = T\partial_T S$. We start at a high temperature with hard-wall
boundary conditions, which has a limiting form for the high-temperature entropy and then lower the temperature in steps to accumulate the entropy for lower temperatures. One must be careful to properly renormalize the entropy per lattice site for a lattice calculation into an entropy per particle for the trapped system. Since the strong-coupling approximation is excellent at high temperatures, it provides a useful benchmark for the LDA-DMFT and GGA-DMFT approaches Iin the following, we sometimes drop the DMFT when we describe the LDA-DMFT and GGA-DMFT approximations, for simplicity).

Because these solution methods are highly optimized, they allow us to directly solve the problem on lattices with 8 million lattice sites and up to 300,000 particles just like in the experimental systems.
The scaling of the code with problem size is discussed elsewhere.~\cite{hpc}

\section{Results} 

\begin{figure*}
\centerline{\includegraphics [width=2.3in, angle=0, clip=on]  {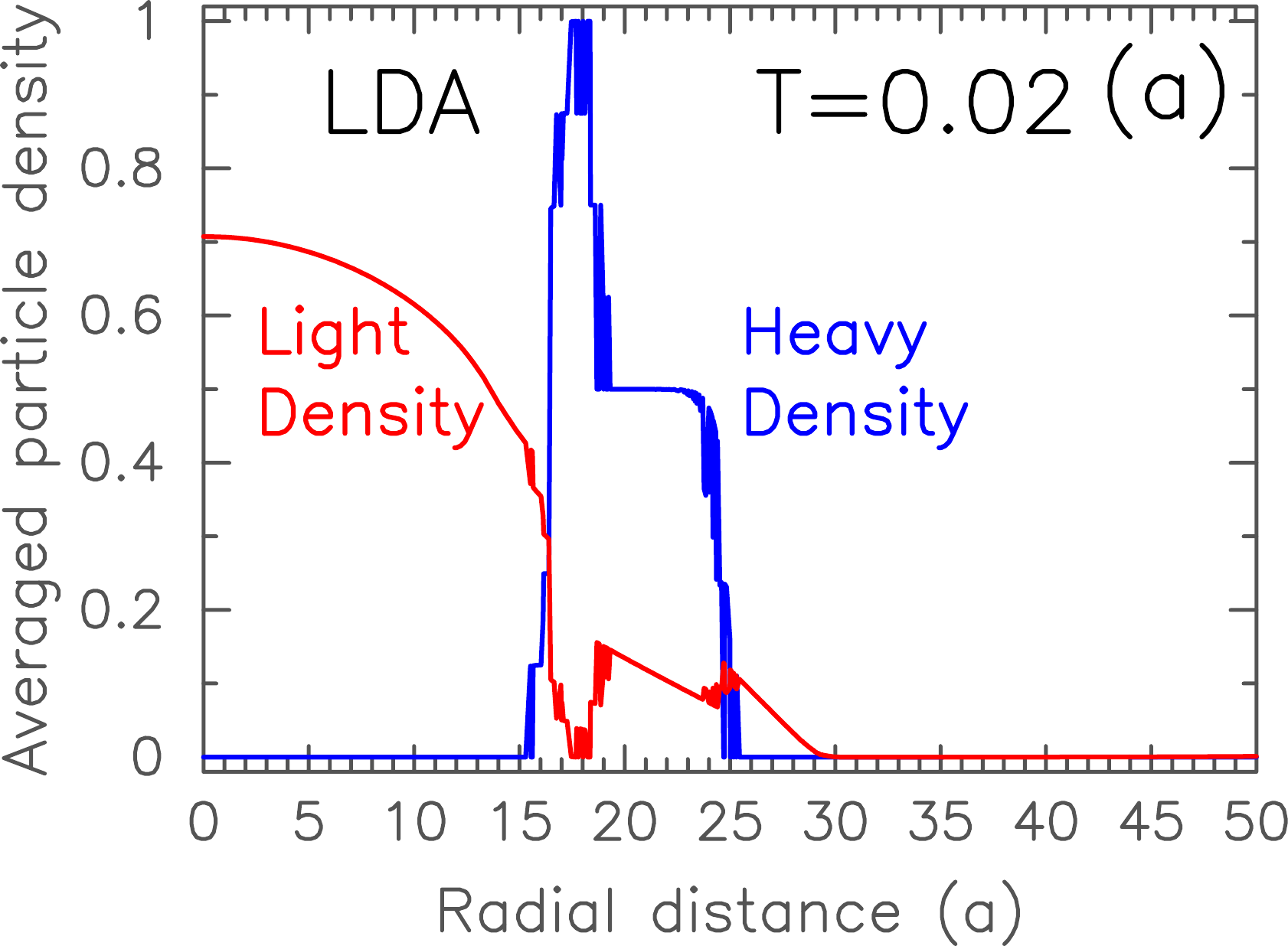}
\includegraphics [width=2.3in, angle=0, clip=on]  {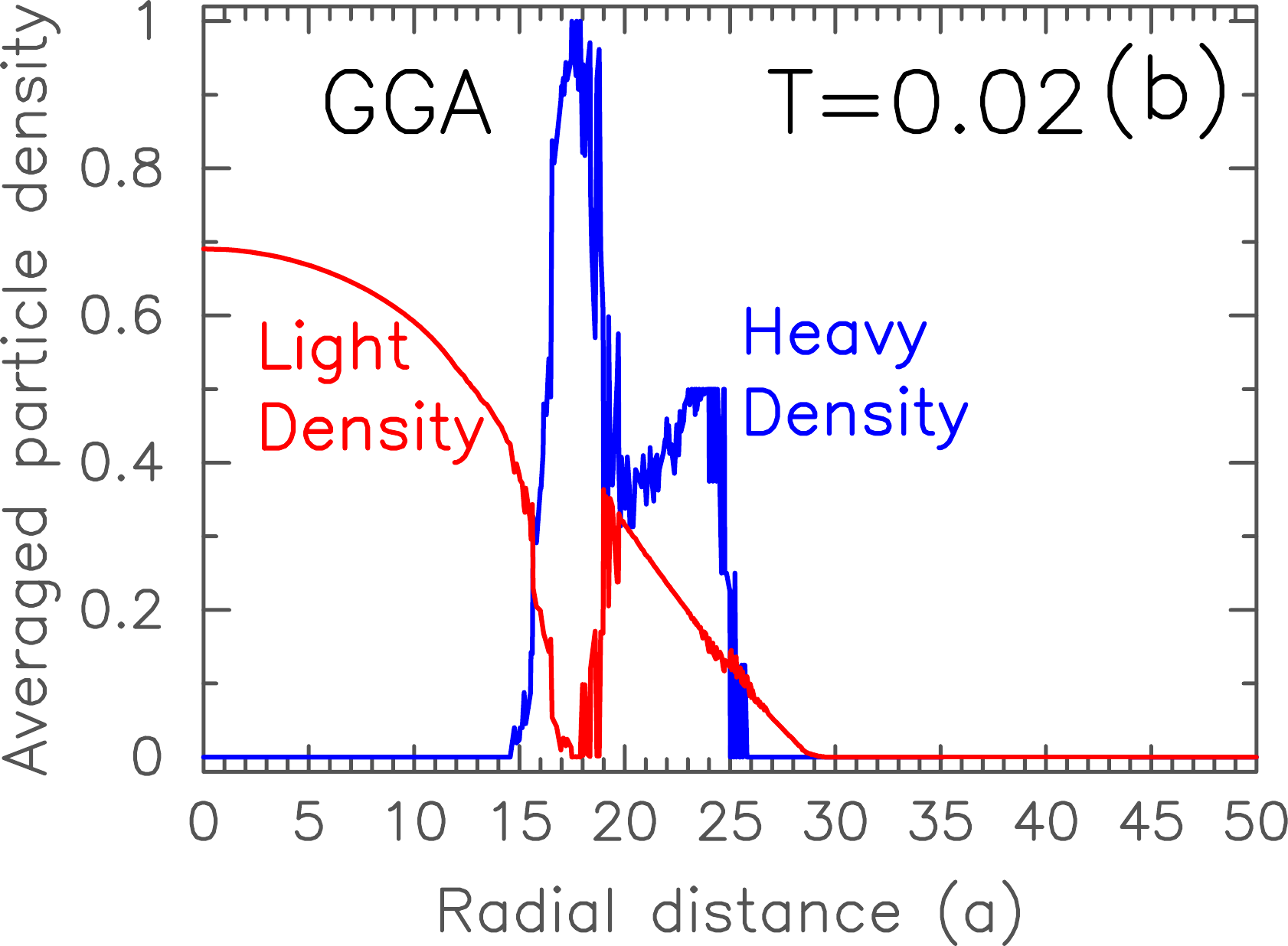}
\includegraphics [width=2.3in, angle=0, clip=on]  {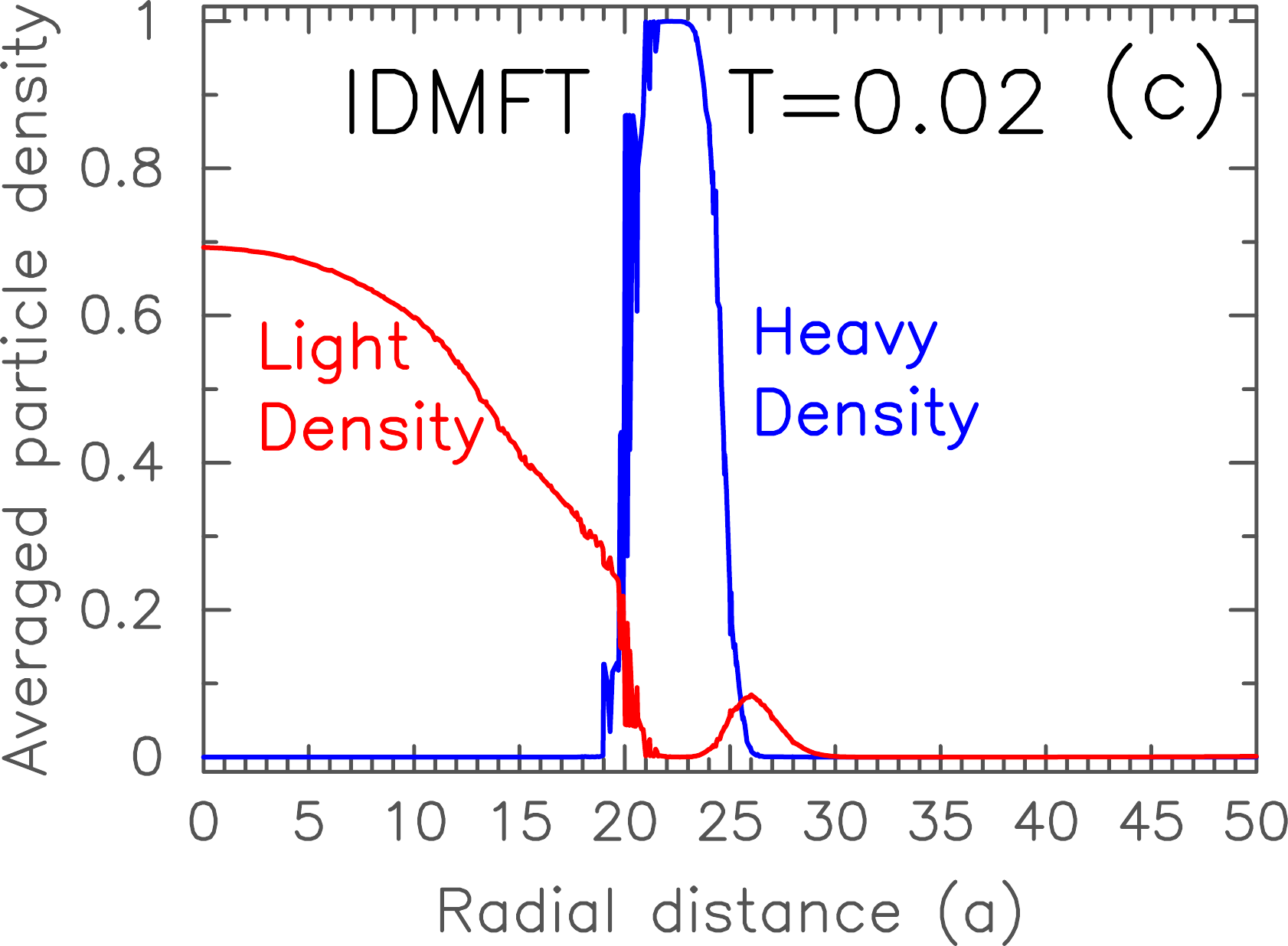}
}
\centerline{\includegraphics [width=2.3in, angle=0, clip=on]  {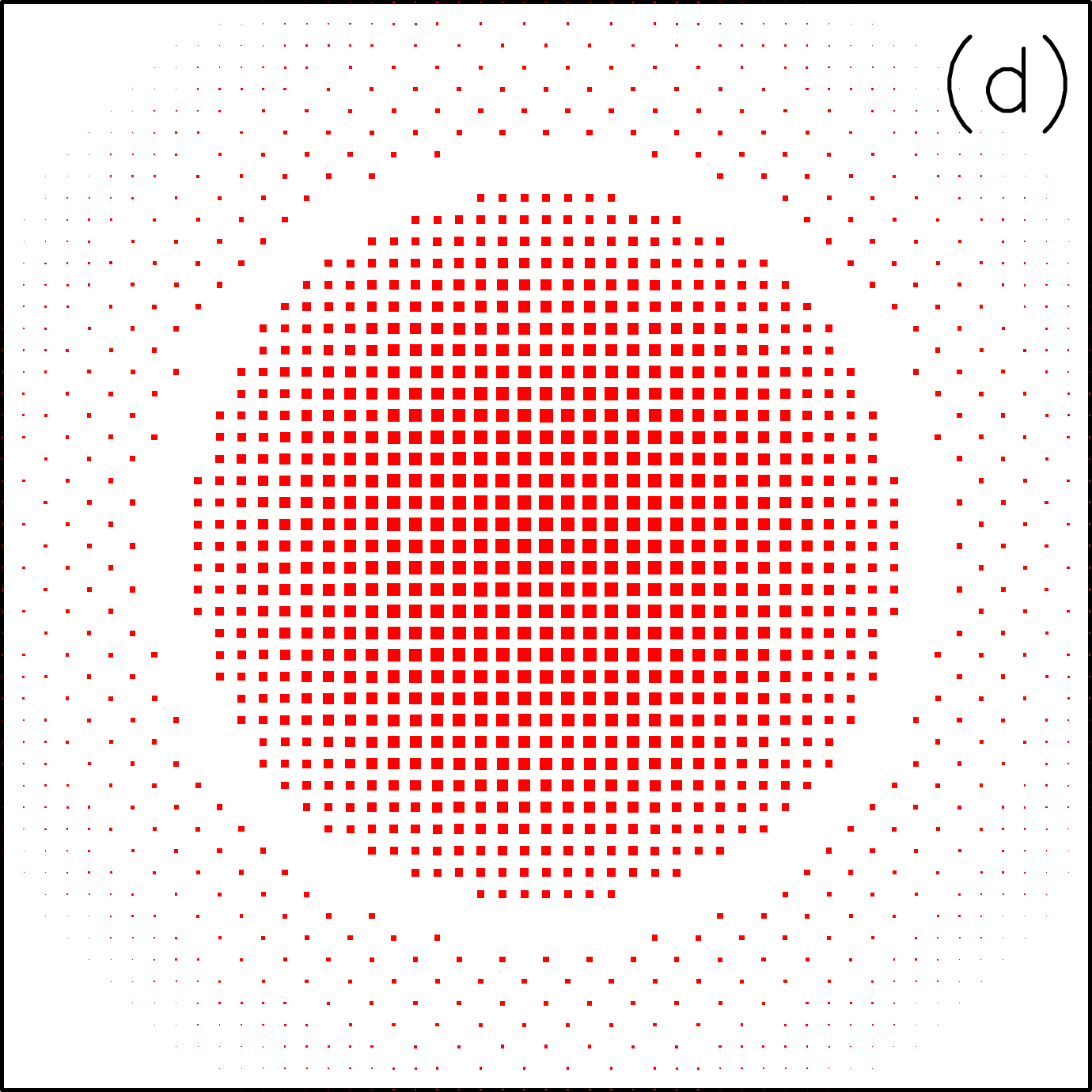}
\includegraphics [width=2.3in, angle=0, clip=on]  {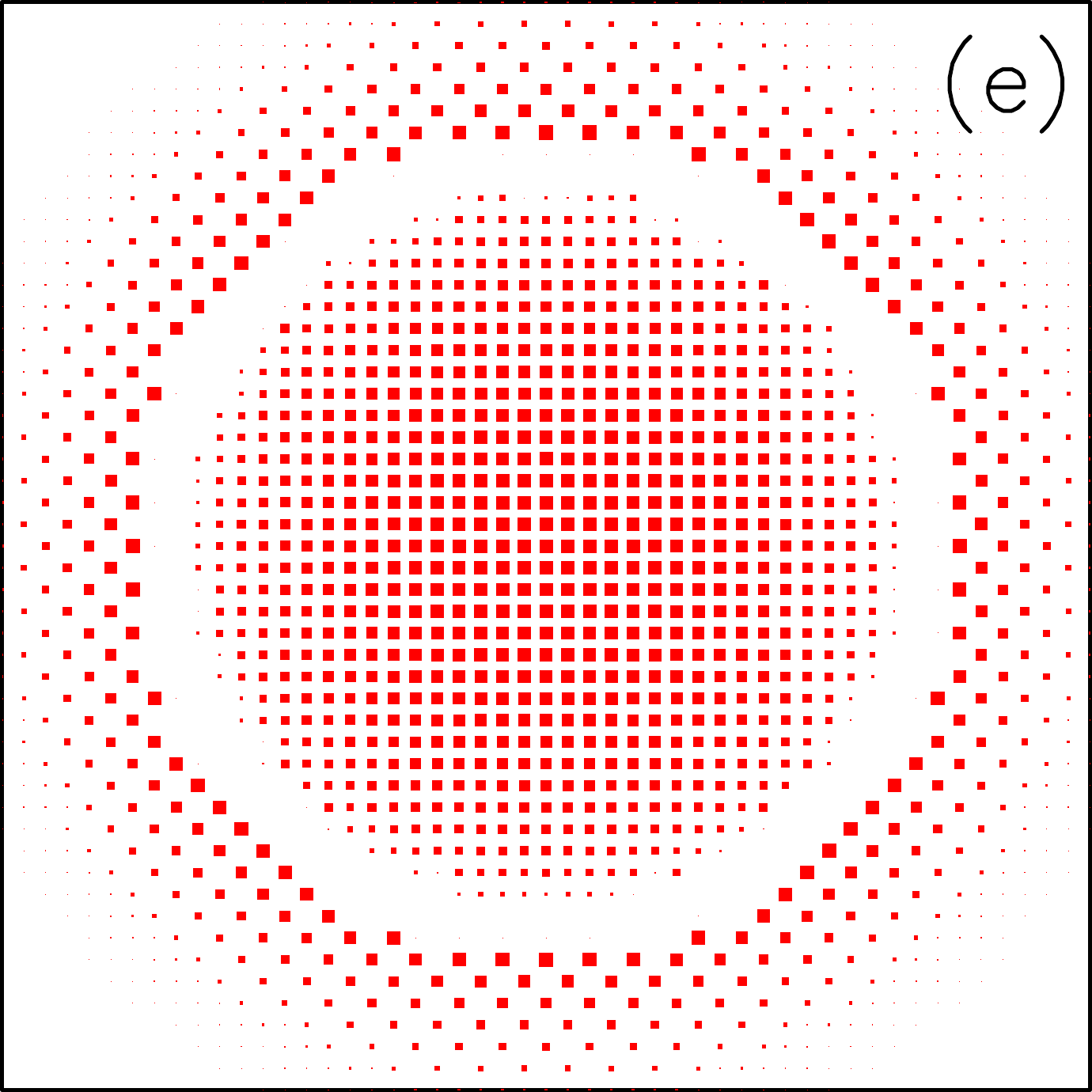}
\includegraphics [width=2.3in, angle=0, clip=on]  {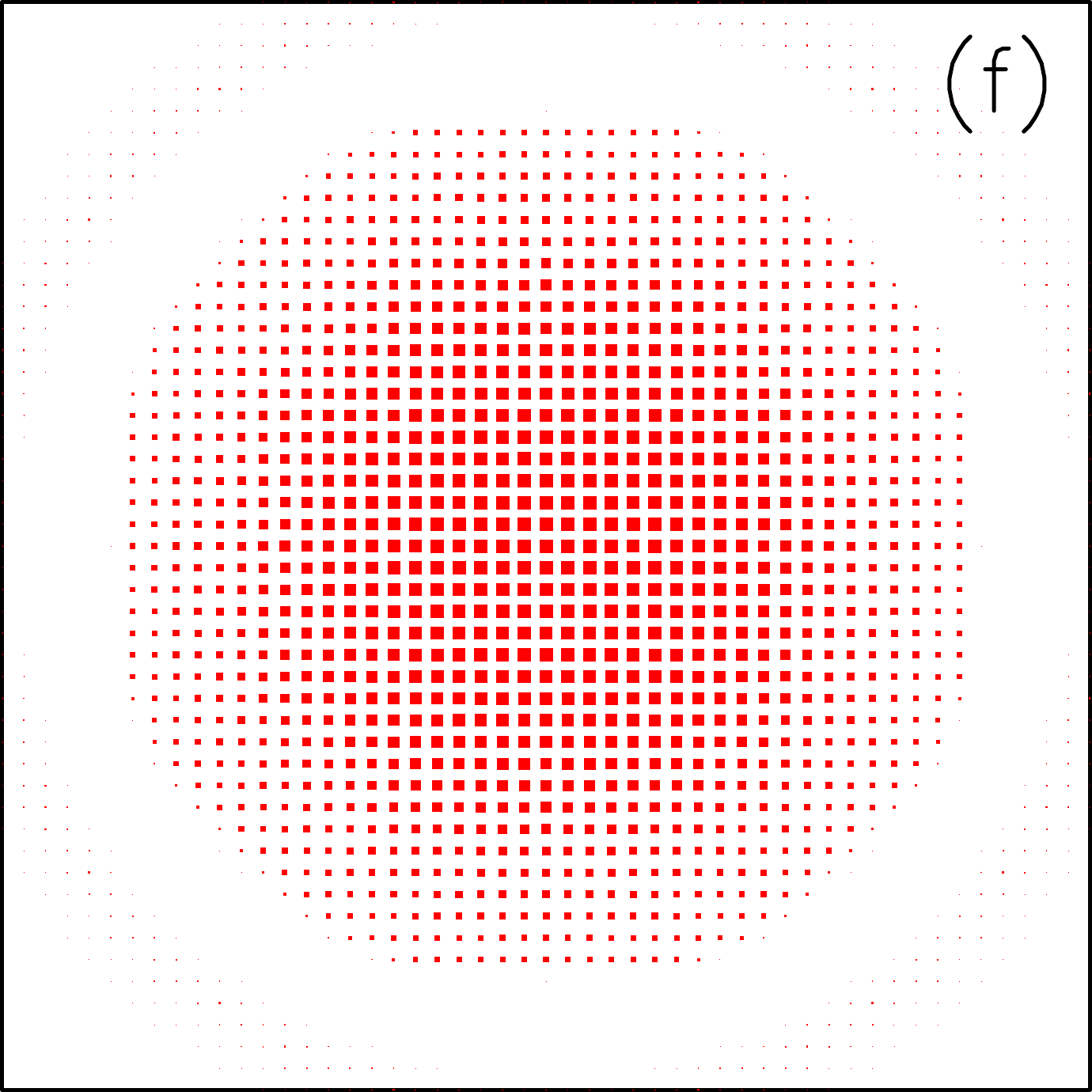}
}
\centerline{\includegraphics [width=2.3in, angle=0, clip=on]  {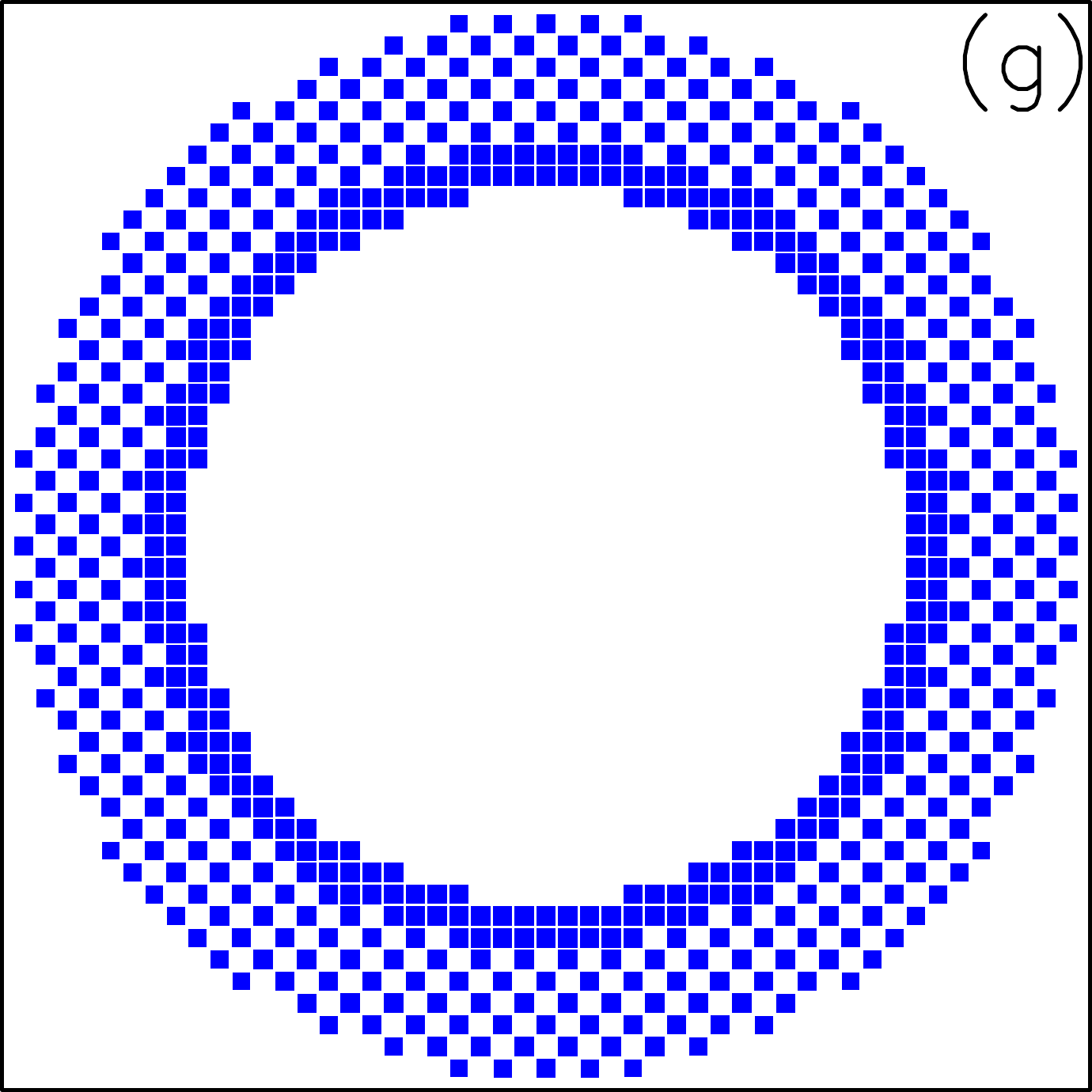}
\includegraphics [width=2.3in, angle=0, clip=on]  {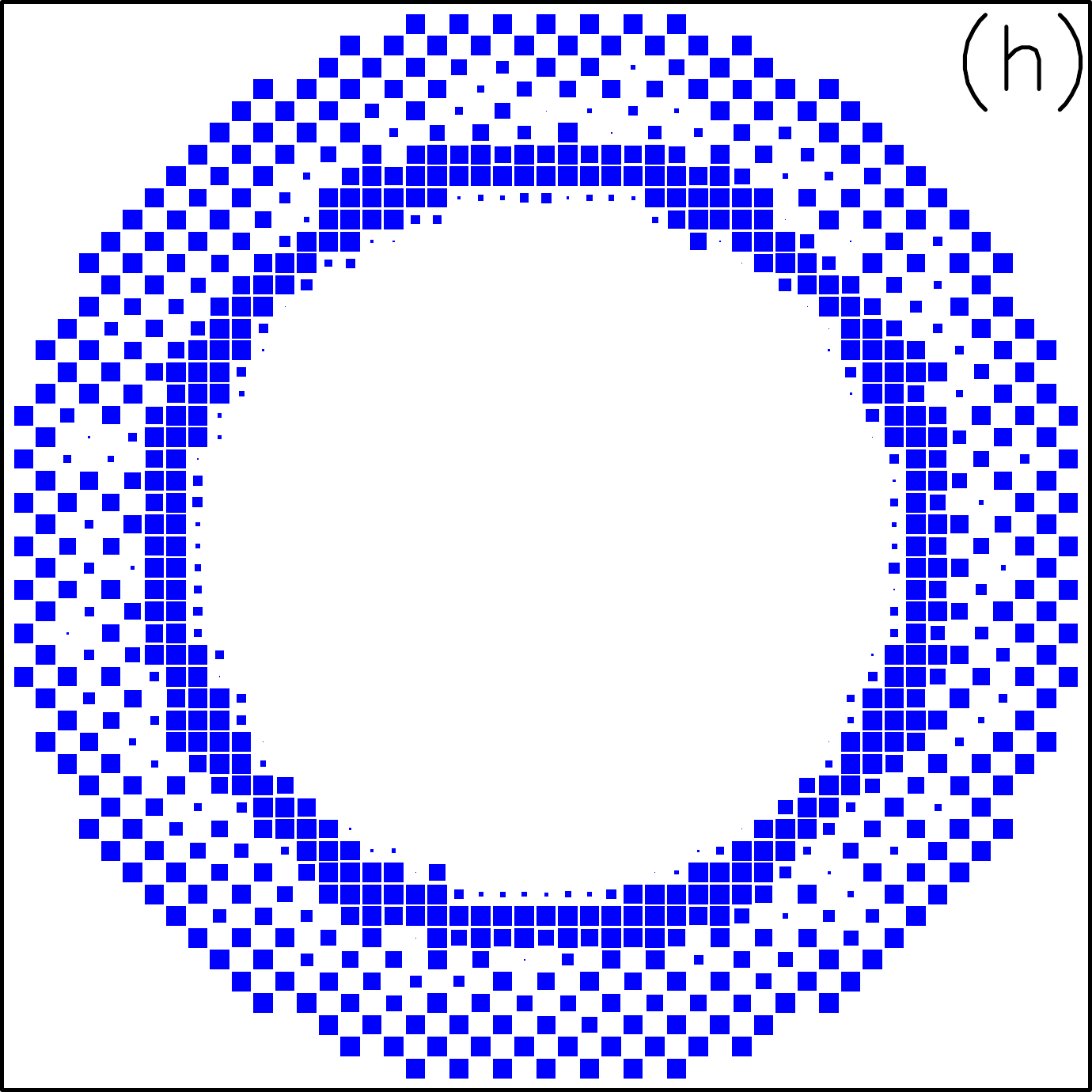}
\includegraphics [width=2.3in, angle=0, clip=on]  {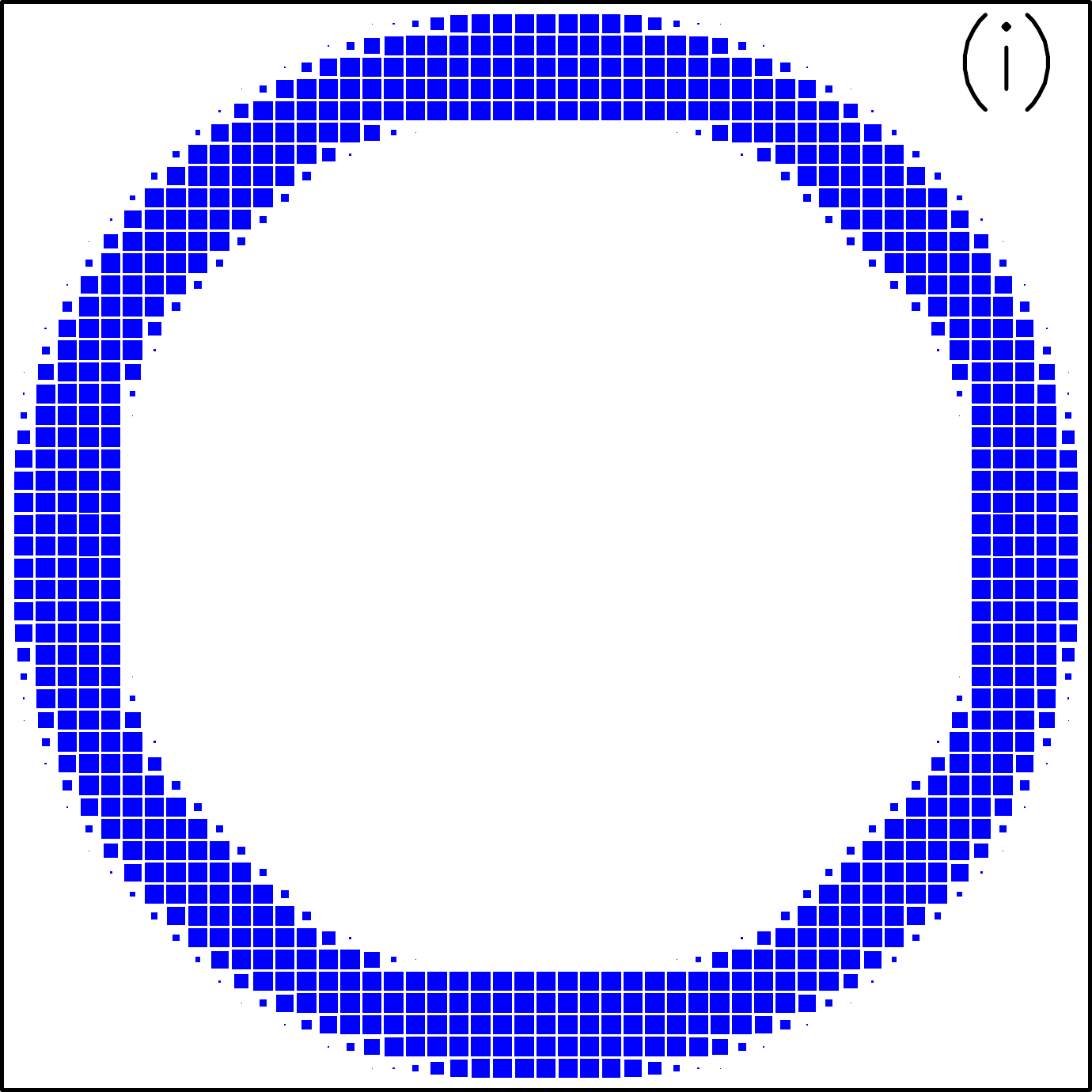}
}
\caption[]{
(Color on-line.) (Upper row) Radial density for T=0.02. Blue is for heavy particles, red is for light particles. From left to right, we have the (a) LDA, (b) GGA, and (c) IDMFT results.
(Middle row) Light particle density for T=0.02 (red). The size of the symbol is proportional to the density of the light particles at that site. From left to right, we have the (d) LDA, (e) GGA, and (f) IDMFT results.
(Lower row) Heavy particle density for T=0.02 (blue). The size of the symbol is proportional to the density of the heavy particles at that site. From left to right, we have the (g) LDA, (h) GGA, and (i) IDMFT results.
\label{fig: t=0.02}
}
\end{figure*}

\subsection{Falicov-Kimball model}

We perform a series of test calculations on the Falicov-Kimball model to examine the robustness of
the LDA-DMFT and GGA-DMFT approaches. The system we chose to examine is a two-dimensional lattice with $|\Lambda|=101\times 101=10,201$. This system is large enough that the density of particles approaches zero at the boundary at high temperature where $T/t=0.5$, and stays within the boundary for all lower $T$. We have 625 light and 625 heavy particles and choose $U_{FK}/t=5$, so we can compare to work done previously~\cite{maska}. The trap potential
is written as 
\begin{equation}
V_i^{(f)}= t\left (\frac{\hbar\omega^{(f)}}{2t}\right )^2\frac{|{\bf R}_i|^2}{ a^2}
\end{equation}
where the $f$ superscript is used for the heavy particles and $\omega$ denotes the trap frequency. Here, we define $\alpha^{(f)}=2t/\hbar\omega^{(f)}$ as the trap parameter, which is set equal to 12.9 for the light particles and 30 for the heavy particles. At low temperatures, this system shows phase separation,
where the light particles concentrate at the center and the heavy particles concentrate in a ring on the periphery of
the system. Because this system has a homogeneous order parameter in the phase-separated state,
it should be  describable by both the LDA and the GGA more accurately than any other phase transition that partially breaks translational symmetry. Hence, it is a useful test case to examine. Note that energies are in units of $t$ and entropy per particle in units of $k_B$ when not explicitly included.

We begin by showing radial profiles for the LDA, the GGA, and the IDMFT results at the same temperature. In Fig.~\ref{fig: highT}, we plot the radial profiles for the light particles (red) and the heavy particles (blue) for $T=0.5$ (upper) and $T=0.15$ (lower). Across each row, we show the (a and d) LDA, the (b and e) GGA, and the (c and f) full IDMFT results. In all the cases, the difference between these curves is difficult 
to see with the naked eye. Indeed, in the normal state, before phase separation sets in, all results agree perfectly, indicating the accuracy of the LDA, and the fact that corrections are very small.

On the other hand, as the temperature is lowered further, the system starts to enter the phase-separated
state, and the bulk systems also see density-wave ordered phases, the most prominent being the 
checkerboard phase. In our results, we can see these effects clearly. We start to see disagreements between the three methods and even the prediction of the wrong ordered phases!

Start with $T=0.1$, shown in Fig.~\ref{fig: t=0.1}, where we first see the effects of the phase separation.
In the top row, we plot the radial density profiles for the three different approximations. Unlike at higher temperature, where the curves were all essentially identical, we see disagreement between all three
methods here. The fast oscillations in the radial profile for the GGA case, occur due to an ordering in the checkerboard
phase at the edge of the region where the light particles give way to the heavy particles (see middle and lower row). Note that
the GGA also shows the beginnings of a faceted stable region instead of a circular region. Of course,
the IDMFT solution doesn't show these features, so they are an artifact of the GGA solution. 

It is easy to understand what is happening here. Both the LDA and the GGA are predicting too tight a compression of the heavy particles around the light ones. The result is that the transition between the two occurs too soon, and ends up lying very close to the regions where the checkerboard phase is stable near half-filling. For the LDA, it doesn't quite look like the checkerboard is stabilized (although there might be a small modulation of the charge), while the GGA has significant regions with obvious checkerboard order. Because the IDMFT solution is somewhat less dense, it stays away from simultaneous half-filling for both particles, and hence does not display the density-wave order.

To understand this behavior further, we lower the temperature even more.  The results at $T=0.05$ are
shown in Fig.~\ref{fig: t=0.05}. Surprisingly, here we see the checkerboard density-wave ordering is far reduced, and appears only on the boundary regions of the heavy particle rings. Looking closely at the LDA  solution, we can see the beginnings of some organization of facets, with the flat regions in the inner and outer
circles for the heavy particle distributions. In the GGA, the behavior is somewhat reduced. It is absent in the IDMFT solution. One might ask, how can such facets form at all in the LDA or GGA, since the potential profiles have circular symmetry, but a little thought shows that if there is any kind of density-wave order, then one might not be able to fit the pattern of the density wave into a circular shape, and hence it can slightly alter the shape of the density distributions, yielding behavior that resembles facets. Of course, the IDMFT can naturally pick up facets if they are energetically favorable, which they clearly are not at this temperature.

Finally, we go to the lowest temperature studied here $T=0.02$ in Fig.~\ref{fig: t=0.02}. Here the LDA shows an innermost phase separated region containing light particles, surrounded by a phase-separated ring containing heavy ones, followed by a checkerboard region, where both species interpenetrate
before the density gets small at large distances. The GGA appears to have a small amount of disorder in the checkerboard phase, which perhaps could be inferred as an indication that it is trying to destabilize the checkerboard phase. But otherwise, it resembles the LDA results quite closely, with the exception that it shows more faceting, as one might expect because it has some sensitivity to nonlocal ordering effects. The IDMFT results, on the other hand, continue to show only phase separation, and they now start to build up facets as one might expect at low temperatures.

Hence, the results we see for the Falicov-Kimball model, which represents mixtures of atoms with different hoppings on an optical lattice (one much bigger than the other), show that at high temperature, the LDA is an excellent approximation, but it rapidly breaks down as one lowers the temperature and enters the ordered phases. This occurs, primarily, because neither the LDA nor the GGA can enforce global constraints and hence favor density-wave ordering much more strongly than it occurs in the real system. 
The IDMFT results, on the other hand, evolve smoothly and, because they invoke global constraints, they show less density-wave ordering patterns. One can also see the stability of the half-filled checkerboard phase in the bulk is so strong that it falsely introduces that phase into the LDA and GGA results, even though they are not present in the IDMFT results. There is a delicate balance between local, or nearly local stability and global stability effects that determines the correct phases in the trap, and these effects require more complex algorithms that properly impose the global constraints, rather than techniques that only sense small variations on the scale of the nearest neighbors.

\subsection{Hubbard model}

We next turn to the Hubbard model, where we apply our results to the experimental results of Refs.~\onlinecite{swiss1} and \onlinecite{swiss2}.
In that work, a strong-coupling perturbation theory was employed to determine the entropy per particle of
 fermion systems with different interaction strengths and different particle densities (the assumption was that under an adiabatic turning on of the optical lattice, the entropy per particle should be the same for all particle densities). Furthermore, it was shown that the entropy per particle appeared to be nearly constant and
fell into a reasonable range in between that of the initial fermionic gas (prior to turning on of the lattice) and the final fermionic gas (after dropping the lattice), where it heated during the experiment due to diabatic and other effects. In addition, all but the lowest $U$ cases worked very well for determining a single entropy per particle for the range of different particle densities. But for the smallest interactions, a case where the strong-coupling perturbation theory is expected to be least accurate, and particularly for low densities, one could see significant deviations of the plots of the double occupancy versus particle number.

We investigate this problem further here by comparing the strong-coupling approach to a more accurate but approximate IDMFT calculation based on the LDA and GGA approximations [since the system is too large to be studied with the IDMFT because it requires about ten million lattice sites for the highest number of fermions (about 300,000)]. But because the bulk DMFT solution is rather fast with the continuous time quantum Monte Carlo algorithm, it is feasible to do the LDA-DMFT and GGA-DMFT calculations on a high-performance computer. We took the parameters for the system from experiment. The mass of the K$^{40}$ atom is 39.964~a.u. and the lattice spacing is 532~nm. Energies are measured in terms of the recoil energy $E_r$ for a potassium atom in a 1064~nm laser. Using a lattice depth of $7E_r$, the hopping is 174~Hz, and the trap frequencies for the $x$, $y$, and $z$, axes are 54.1~Hz, 50.1~Hz, and 129.1~Hz, respectively. We use these precise trap frequencies, which create a slightly anisotropic trap. The trap potential energy is written as 
\begin{equation}
\frac{1}{2}m\left ( \omega_x^2 x^2+\omega_y y^2+\omega_z^2 z^2\right ) =
\hbar t \left ( \frac{a^2i^2}{R_x^2}+\frac{a^2j^2}{R_y^2}+\frac{a^2k^2}{R_z^2}\right )
\end{equation}
where we introduced the trap lengths $R_\alpha$ given by $R_x=10.3a$, $R_y=11.1a$ and $R_z=4.3a$. We choose the lattice to initially be a cube of 250 lattice sites on each edge. Next, we compute the maximal potential at the point $(0,125a,0)$ which is at the center of the cube face where the potential grows the slowest, and then discard all points in the lattice with a potential energy due to the trap that is larger than this cutoff. The net number of lattice sites used was then 1,964,881. But we reduced these
lattice sites into equivalence classes due to the three reflection planes, which involved 253,230 equivalence classes for different lattice sites (in an irreducible wedge). This is the number of lattice sites for which we needed to solve the impurity problem with the QMC algorithm during each step of the iteration. This process is easy to parallelize despite the fact that at low temperatures the times to obtain solutions for different fillings could vary by more than an order of magnitude. This is done simply by employing a master-slave algorithm and dividing the work according to ``first finished gets a new job''. The code scaled linearly on up to 43,500 cores. About 15 million cpu hours were required to run these codes on a Cray XE6.~\cite{hpc}

In the experiment,  systems of up to 300,000 particles were cooled down to an entropy per particle of about $1.3~k_B$. Two different hyperfine states of K were employed for the spin-up and spin-down states of the Hubbard model. The optical lattice was then turned on, experiments performed, and then the lattice was dropped. The ending entropy per particle was around $2.5~k_B$. If we assume the entropy is held constant during the experiment in the trap, then we would immediately guess that the entropy per particle is halfway between these two limits or $1.9~k_B$, but there might be some variations depending on the interactions and on whether both the turning on and turning off of the lattice heat the system in the same fashion. Hence, we would like to be able to independently assess the value of the entropy per particle in the optical lattice. This was done by plotting the double occupancy versus the total number of particles and fitting to isoentropic curves for the double occupancy versus the number of particles. The latter were found via a strong coupling approach. Excellent results were found for $U=24.6t$, $19.2t$, and $14.4t$, but when the case with $U=8.4t$ was examined, there were clear deviations between the experimental data and the isoentropic curves at low particle numbers. Hence, we choose to examine this case with the LDA and GGA approaches employing a weak-coupling continuous-time QMC solver instead of the strong-coupling solver.

Numerically, we achieve this by picking a fixed number of particles and setting $U=8.4t$. Then we start at high temperature and use the LDA or GGA algorithms to compute the total number of particles for a given chemical potential. This is then adjusted until the total number of particles agrees with the target value; because the LDA and GGA results are so close here, we use the same global chemical potential for both, which results in a slightly different total number of particles for each. Then we extract the double occupancy and the total energy. By using the integral formula for the entropy, we also determine that for each temperature. We then reduce the temperature and repeat. In our calculations, we typically used about 12 different temperatures to establish a fine grid of entropy versus $T$, fraction of doubly occupied particles $d$ versus $T$, and $\mu(T)$. This is then repeated for the different experimental densities (15 different total number of particles chosen to match experiment), and the data is then interpolated to determine the isoentropic curves. These curves are then compared to the experimental data. As an example, we show a series of calculations for the target particle number $N=174,518$ in Tables~\ref{table: lda} and \ref{table: gga}. One can see that the two calculations are quite close to each other, but they deviate rather significantly from the strong-coupling
approach at low temperatures.

\begin{table}[htb]
\begin{tabular}{ r | r c c c c c }
$T/t$ & $\mu$~~~ & $N/N_0$ & $d$ & $E/N$ & $S^{LDA}/N$ & $S^{strong}/N$\\
\hline
20.0 & -17.3833 & 23.518 & 0.0809 & 30.542 & 3.474 & 3.474\\
15.0 & -5.4714 & 23.536 & 0.1045 & 25.177 & 3.161 & 3.160\\
10.0 & 4.6824 & 23.555 & 0.1477 & 19.139 & 2.658 & 2.658\\
7.0   & 9.5722 & 23.556 & 0.1918 & 15.655 & 2.235 & 2.231\\
5.0  & 12.1948 & 23.576 & 0.2356 & 13.621 & 1.886 & 1.869\\
3.3  & 13.9113 & 23.591 & 0.2884 & 12.227 & 1.538 & 1.493\\
2.5 &  14.5775 & 23.614 & 0.3242 & 11.685 & 1.348 & 1.267\\
2.0 &  14.9092 & 23.609 & 0.3507 & 11.341 & 1.193 & 1.112\\
1.5 &  15.1753 & 23.620 & 0.3807 & 11.097 & 1.051 & 0.948\\
1.0 &  15.3812 & 23.636 & 0.4156 & 10.914 & 0.898 & 0.765\\                                                                                                                                                            
0.8 &  15.4460 & 23.642 & 0.4311 & 10.863 & 0.841 & 0.686\\
\hline
\end{tabular}
\caption{Data employed in the LDA calculation for $N=174,518$. $N_0=7,393$ is the characteristic particle number, $d$ is the fraction of doubly occupied particles, $E/N$ is the total energy per particle in units of $t$, and $S/N$
is the entropy per particle in units of $k_B$ for the LDA and strong-coupling approximations. We have $U/t=8.4$. Note how the strong-coupling
approximation agrees well with the LDA-DMFT-QMC calculation until the temperature becomes low enough that they start to deviate.}
\label{table: lda}
\end{table}

\begin{table}[htb]
\begin{tabular}{ r | r c c c c c }
$T/t$ & $\mu$~~~ & $N/N_0$ & $d$ & $E/N$ & $S^{GGA}/N$ & $S^{strong}/N$\\
\hline
20.0 & -17.3833 & 23.529 & 0.0808 & 30.555 & 3.474 & 3.474\\
15.0 & -5.4714 & 23.545 & 0.1044 & 25.186 & 3.161 & 3.160\\
10.0 & 4.6824 & 23.561 & 0.1477 & 19.139 & 2.657 & 2.658\\
7.0   & 9.5722 & 23.572 & 0.1917 & 15.656 & 2.234 & 2.231\\
5.0  & 12.1948 & 23.580 & 0.2355 & 13.622 & 1.885 & 1.869\\
3.3  & 13.9113 & 23.598 & 0.2882 & 12.230 & 1.537 & 1.493\\
2.5 &  14.5775 & 23.617 & 0.3241 & 11.686 & 1.347 & 1.267\\
2.0 &  14.9092 & 23.607 & 0.3507 & 11.340 & 1.191 & 1.112\\
1.5 &  15.1753 & 23.619 & 0.3808 & 11.097 & 1.049 & 0.948\\
1.0 &  15.3812 & 23.633 & 0.4158 & 10.913 & 0.897 & 0.765\\                                                                                                                                                            
0.8 &  15.4460 & 23.637 & 0.4314 & 10.861 & 0.838 & 0.686\\
\hline
\end{tabular}
\caption{Data employed in the GGA calculation for $N=174,518$. $N_0=7,393$ is the characteristic particle number, $d$ is the fraction of doubly occupied particles, $E/N$ is the total energy per particle in units of $t$, and $S/N$
is the entropy per particle in units of $k_B$ for the LDA and strong-coupling approximations. We have $U/t=8.4$. Note how the GGA and LDA results are nearly identical.}
\label{table: gga}
\end{table}

Since the strong-coupling calculation is much more efficient, we compare our results to the strong-coupling results. We found that the entropy versus $T$ for the GGA solution starts to deviate from strong-coupling results (with the DMFT results showing a higher entropy per particle than the strong-coupling results) at temperatures in the range of $2.5t\le T\le 3.3t$ for both high and low densities of particles (see Fig.~5 of Ref.~\onlinecite{hpc} for a plot and look at the two tables). This is a little higher than the rule of thumb that the strong-coupling approach fails once the temperature reaches around $t$. Unfortunately, because the DMFT solutions are known to overestimate the entropy at low temperatures, it is not clear which result is more accurate in the
temperature range from about $3t$ to $0.5t$. We operate under the assumption that the DMFT results are better here because the interaction strength is not particularly strong.

\begin{figure}
\centerline{\includegraphics [width=3.0in, angle=0, clip=on]  {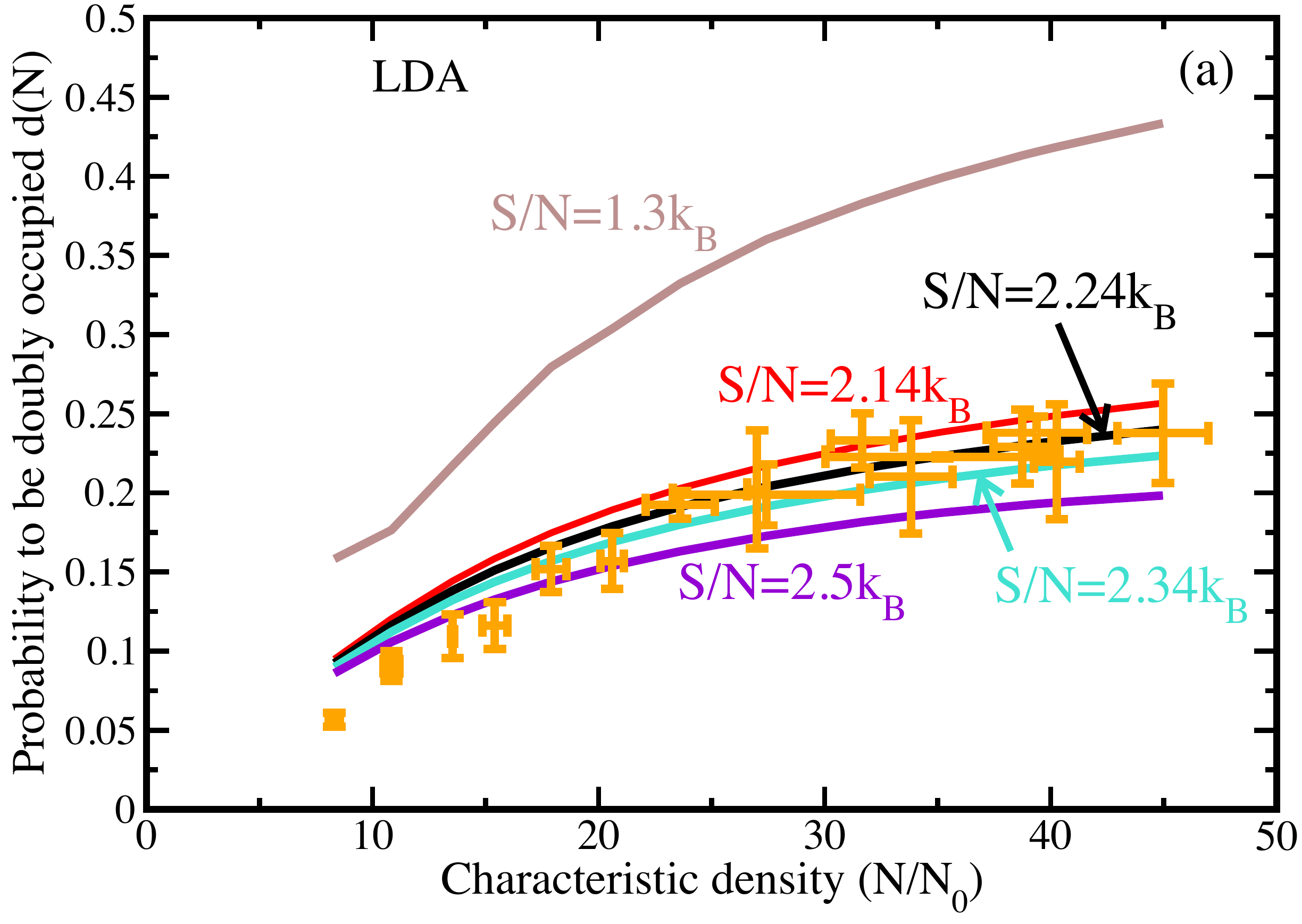}
}
\centerline{\includegraphics [width=3.0in, angle=0, clip=on]  {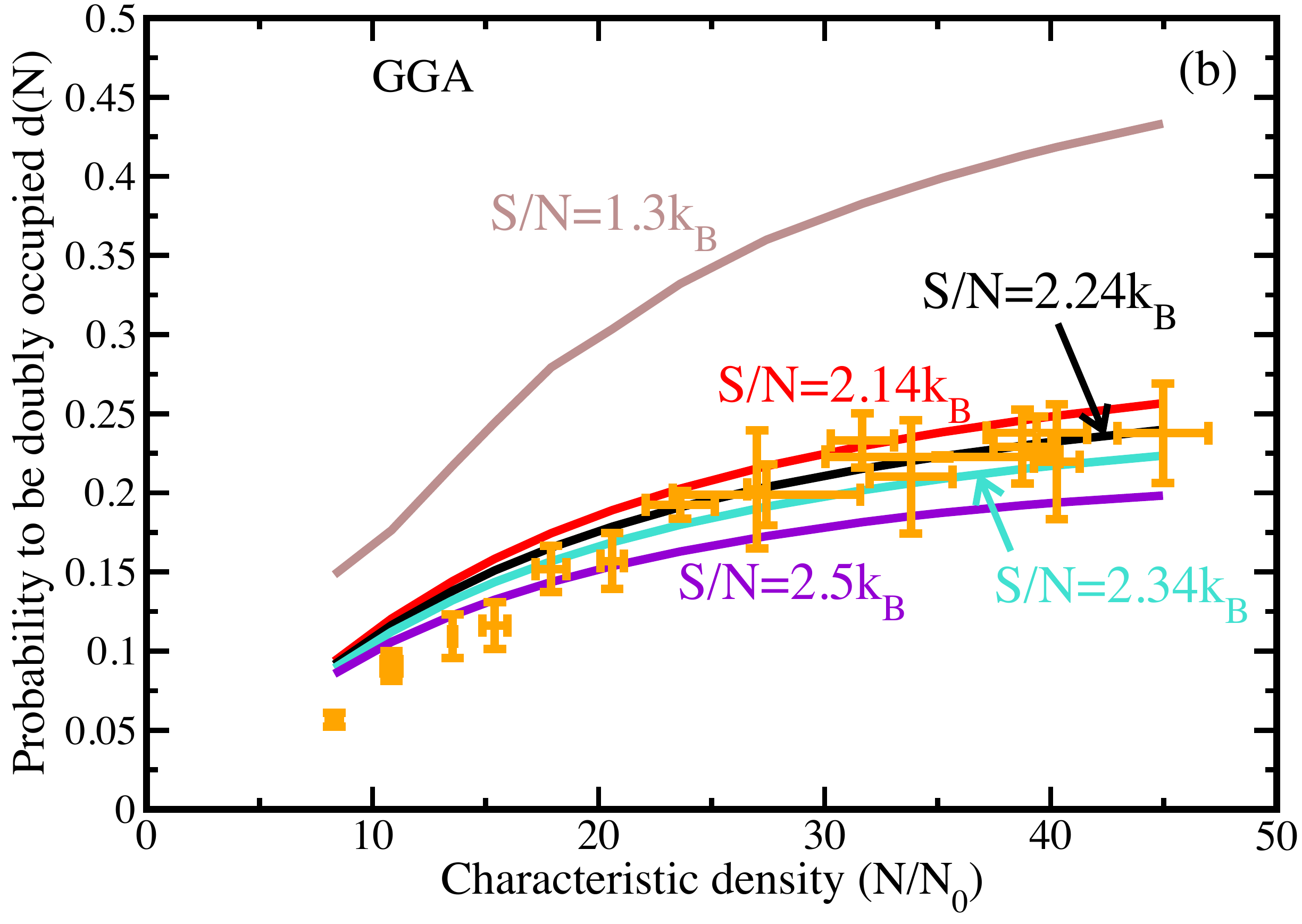}
}
\caption[]{
(Color on-line.) Double occupancy versus particle number for experiment (symbols with error bars) and
isoentropic lines with the entropy per particle equal to the most-likely fit interval 2.14, 2.24, or 2.34 $k_B$ and the extreme values at 1.3 and 2.5~$k_B$ for (a) LDA and (b) GGA. In both cases, we have $U/t=8.4$.
\label{fig: hubbard}
}
\end{figure}

The fitting of the data is then rather straightforward. We plot a series of isoentropic curves for the double occupancy versus the total number of particles and compare to the data and to the data shifted upward and downward by the standard deviation of the double occupancy at each point (see Fig.~\ref{fig: hubbard}). Then we can estimate both by eye and by evaluating a weighted least-squares fit what the optimal entropy per particle is.  Note that the low particle number data still do not fit so well. Hence the use of LDA or GGA does not significantly improve the analysis. 

\begin{figure}
\centerline{\includegraphics [width=3.0in, angle=0, clip=on]  {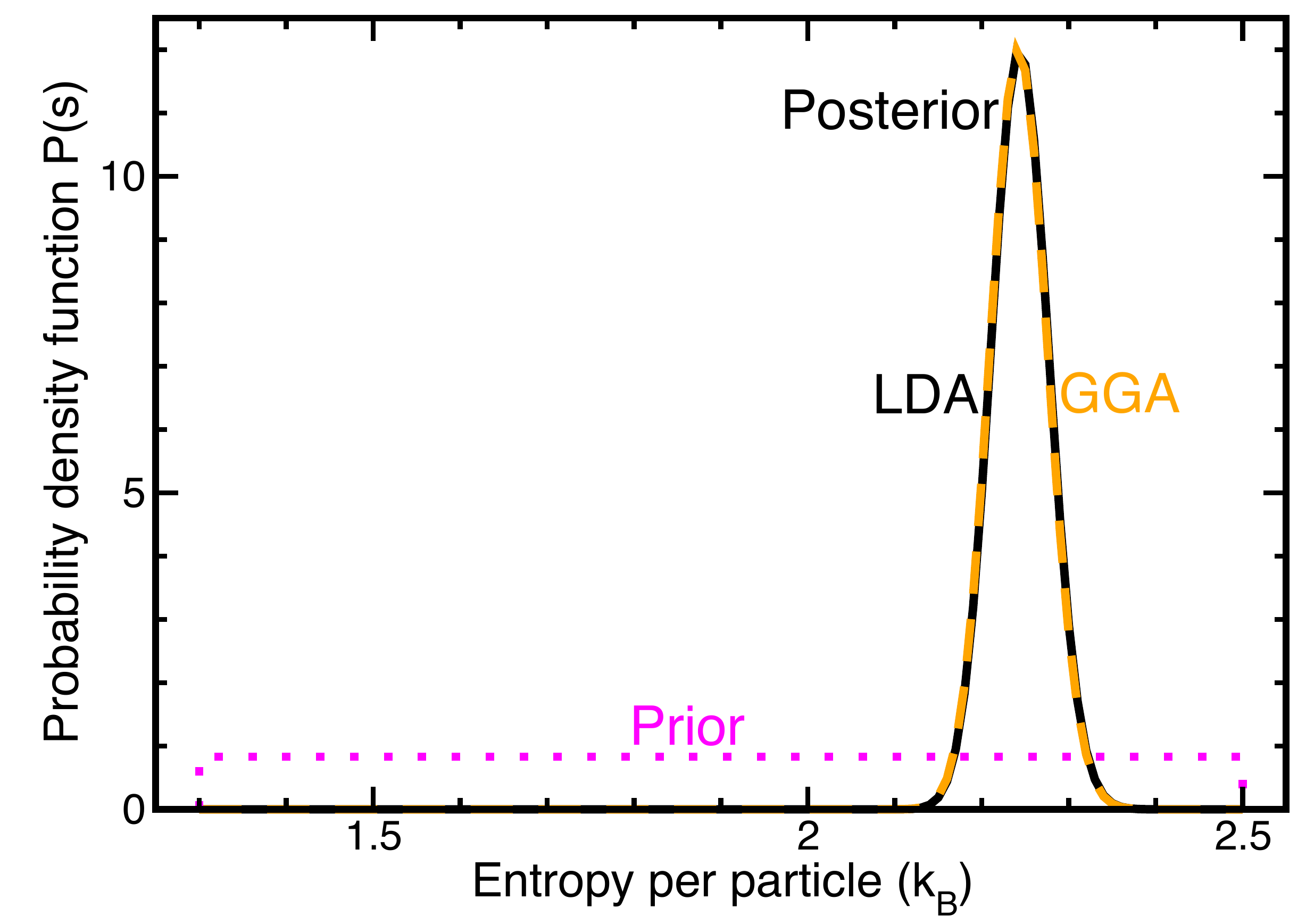}
}
\caption[]{
(Color on-line.) Posterior PDF for the LDA (black, solid) and GGA (orange, dashed) approaches, compared
to the prior PDF (magenta, dashed). Note how it becomes strongly peaked around 2.24~$k_B$.
\label{fig: pdf}
}
\end{figure}

To be more quantitative, we use a Bayesian inference approach to determine the posterior
probability density function (PDF)  as determined by the experimental measurements. We start from a prior
PDF that is flat between the initial experimental value for the entropy per particle 1.3~$k_B$ and the final value 2.5~$k_B$ (since we only know the possible range for the entropy and all possibilities are {\it a priori} equally likely). For each experimental measurement, we compare to a theoretical estimate which results from the QMC calculations within the LDA or GGA approach, which are extended from their computed values by a quadratic Lagrange interpolation formula. Then we form the chi-square
\begin{equation}
\chi^2(s)=\sum_i \left [ \frac{d_i^{exp}(N_i)-d_i^{th}(s,N_i)}{\sigma_i}\right ]^2.
\end{equation}
Here, $d_i^{exp}(N_i)$ is the measured experimental probability for an atom to be paired when there are on average $N_i$ atoms in the system, $d_i^{th}(s,N_i)$ is the theoretical prediction for the probability to be paired
at a fixed value for the entropy per particle; this is calculated by interpolating the QMC data. The symbol $\sigma_i$ is the experimental standard deviation for each experimental measurement at $N_i$ of the pairing probability. 
The posterior probability for the system then follows from Bayes rule: $P(s|d)=P(s)P(d|s)/P(d)\propto P(s)\exp[-\chi^2(s)/2]$. The posterior PDF is the product of the prior PDF (which is chosen to be uniform) and the exponential of minus one half the chi-square. The posterior PDF is plotted in Fig.~\ref{fig: pdf} for the LDA (black solid) and for the GGA (orange dashed). One can see that they nearly lie on top of each other. The simplest estimate for the entropy is the maximum {\it a posteriori} (MAP) estimate, which is the entropy per particle at the maximum of the PDF, which is 2.243~$k_B$ for the LDA and 2.242~$k_B$ for the GGA. We also can calculate the average value of the entropy per particle, by integrating over the PDF along with the variance, which is called the least mean squares (LMS) estimate. It yields $s=2.244\pm 0.033~k_B$ for the LDA and $s=2.243\pm 0.033~k_B$ for the GGA; the strong-coupling fit was $s=2.25~k_B$. The reason why the LDA and GGA approaches do not improve things very much is that the fit temperature 
is rather high and hence is in the regime where the LDA, GGA, and strong-coupling approaches all agree
quite closely.
In particular, we find that the GGA and LDA results are so close for this case that they essentially lie on top of each other. This is not too surprising since we are rather far from the temperature where the system will start to order. Unfortunately, we are not able to quantitatively answer the question as to whether this LDA or GGA analysis is more accurate than the strong-coupling analysis, because we had no unbiased way to choose between them.

One additional point to note, however, is that the lowest point in each figure for $S/N=1.3$~$k_B$ in Fig.~\ref{fig: hubbard} appears to be a bit higher than we would have expected. This arises because of the problems the DMFT has in overestimating the entropy at low temperature. In particular, at half-filling, in a bulk system, the entropy per lattice site cannot get below $\ln 2$~$k_B$ until the systems orders into an antiferromagnetic phase, which can quench the entropy. We believe that the entropy we calculated at the lowest temperature $T=0.5$ for $N=61,445$, is high and when we extrapolate to find the double occupancy for the $S/N=1.3$~$k_B$ curve, we find
it produces results that are too high. The situation is a little better for the GGA than for the LDA, primarily because the entropy in the GGA is a bit smaller than the entropy for the LDA at $T=0.5$ (1.305 versus 1.338~$k_B$, respectively). For both points, we employed a linear extrapolation using $T=0.7$ and $T=0.5$ data.

\section{Conclusion} 

In this work, we showed how to develop the first correction to the LDA by taking into account gradient terms in the Dyson equation for the Green's function. We found that the algorithm for the GGA within a DMFT approach involved only a small change from the LDA. In the normal state, at temperatures above any of the ordering temperatures, the LDA, GGA, and IDMFT results all agree very well with one another. This result strongly supports the accuracy for using the LDA for such normal state calculations of the properties of ultracold atoms on an optical lattice. 

However, when we enter the ordered phase, the LDA and GGA become inaccurate, often predicting the wrong transition temperature and also predicting the wrong ordered phase. We have so far not found there to be any simple way to fix this, as the problem appears to require properly applying global constraints as opposed to local and nearly local constraints. Hence, we conclude that the LDA is poor for these kinds of
problems when the temperature is low enough that it falls into a regime where the system orders.

This does not rule out the approach as being good to describe systems in nonequilibrium with mass transport. Such a problem is beyond the scope of this article, but there has been some preliminary work on this question.\cite{dirks}

Finally, we conclude that in most cases where the LDA works well, the GGA does not provide a significant improvement, and when the LDA fails, the GGA does not fix the failures, so there does not seem to be a huge need for the GGA {\it per se}, unless it works well at repairing the problem with nonequilibrium mass transport, which should be looked at critically from this perspective. However, calculating results for the GGA is only slightly more complicated than the LDA, and doing so can serve as a useful test on the accuracy of the LDA---in cases where the LDA and GGA agree, it is likely that their results are accurate, in
cases where they disagree, it is likely that neither is accurate.

\section*{Acknowledgments}
J. K. F.  acknowledge support from the National Science Foundation under grant number PHY-1314295. J. K. F. also acknowledges support from the McDevitt bequest at Georgetown University. 
H. R. K. was supported by the Department of Science and Technology (India).
DoD HPC computer time was provided on Cray XE6 machines located at the Arctic Region
Supercomputer Center (ARSC), the US Air Force Engineering and Research and Development Center (AFRL) and the US
Army Engineering Research and Development Center (ERDC). The numerical work was performed in 2010--2011, where this project was supported primarily by a Challenge Project
(DARPA-C4J) and a capabilities application project. During this early phase, the project was supported by
the DARPA optical lattice emulator program, with funds from the Army Research Office.
We thank the Esslinger group for providing us with their experimental
data and for conversations with Niels Strohmaier and Leticia Tarruell that helped clarify the data.

\end{document}